\documentclass[conference]{IEEEtran}
\usepackage{graphicx}
\usepackage[svgnames,table]{xcolor}
  \definecolor{lightgray}{gray}{0.9}
\usepackage{url}
\usepackage{cleveref} 
  \Crefname{section}{Sec.}{Secs.}
  \Crefname{table}{Tab.}{Tabs.}
  \Crefname{figure}{Fig.}{Figs.}
\usepackage{booktabs}
\usepackage{multirow}
\usepackage{tabularx}
\usepackage{wrapfig}
\usepackage[english]{babel}
\usepackage{flushend}
\usepackage{enumerate}
\usepackage{ragged2e}  %
  \newcolumntype{P}[1]{>{\raggedright\arraybackslash}p{#1}}
  \newcolumntype{?}{!{\vrule width 1pt}}
\usepackage{amssymb}
\usepackage{colortbl}
\usepackage{paralist}
\usepackage{subfig}
\usepackage{tikz}
  \usetikzlibrary{arrows,arrows.meta,positioning,automata,
    shapes,decorations.pathreplacing,calc}

\newcommand{\ies}{i.e.,\ }

\newcommand{\egs}{e.g.\ }

\usepackage[colorinlistoftodos]{todonotes}
\presetkeys{todonotes}{inline}{}

\usepackage{ctable}

\IEEEoverridecommandlockouts

\begin{document}
\title{Arguing from Hazard Analysis in Safety Cases:\\
  A Modular Argument Pattern\textsuperscript{*}
  \thanks{\textsuperscript{*}%
    IEEE Reference Format: Gleirscher, M. \&
    Carlan, C. Arguing from Hazard Analysis in Safety Cases: A Modular
    Argument Pattern. High Assurance Systems Engineering (HASE), 18th
    Int. Symp., 2017,
    DOI: 10.1109/HASE.2017.15.
  }
}

\author{
  \IEEEauthorblockN{Mario Gleirscher}
  \IEEEauthorblockA{Technische Universit\"at M\"unchen, Munich,
    Germany\\
    Email: mario.gleirscher@tum.de}
  \and
  \IEEEauthorblockN{Carmen Carlan}
  \IEEEauthorblockA{fortiss GmbH, M\"unchen, Germany\\
    Email: carlan@fortiss.org}
}

\maketitle  
\begin{abstract}
  We observed that safety arguments are prone to stay too abstract,
  \egs solutions refer to large packages, argument strategies to
  complex reasoning steps, contexts and assumptions lack
  traceability. These issues can reduce the confidence we require of
  such arguments.
  In this paper, we investigate the construction of confident
  arguments from (i) hazard analysis (HA) results and (ii) the design
  of safety measures, \ies both used for confidence evaluation.  We
  present an argument pattern integrating three HA techniques, \ies
  FTA, FMEA, and STPA, as well as the reactions on the results of
  these analyses, \ies safety requirements and design increments.  We
  provide an example of how our pattern can help in argument
  construction and discuss steps towards using our pattern in formal
  analysis and computer-assisted construction of safety cases.
\end{abstract}


\section{Introduction}
\label{sec:intro}

We give a short introduction into safety cases, safety arguments, goal
structures, and hazard analysis (HA), point out one important problem we
perceive when building safety cases, and, finally, provide an approach
to solve this problem.

\subsection{Background and Terminology}
\label{sec:safetycases}

According to Bishop and Bloomfield~\cite{Bishop1998}, a \emph{safety
  case}\footnote{The discussion of how our approach relates to the more
  general concepts of \emph{assurance case} and \emph{dependability
    case} is out of scope of this paper.} should comprehensibly convey
a valid argument that a specific system is acceptably safe in a
specific operational context.  Hereby, the \emph{safety argument}
captures the reasoning from basic facts---the evidence---towards the
claims to be established---the safety goals.  Graphs called \emph{goal
  structures} represent and document such an
argument~\cite{Kelly1998}.  To make the process of safety case
construction more systematic, several authors
\cite{Kelly1997,Alexander2007} propose \emph{argument patterns} and
provide a structure for developing lower level patterns.

\emph{Hazard identification} relies on expert knowledge, \egs in
terms of guide words or \emph{defect classifications}, identifying
types of component failures~\cite{Rausand1996,%
Gleirscher2014a},
defects in software processes~\cite{Chillarege1992}, 
accident causal factors~\cite{Leveson2004}, and 
destructive goals for software tests~\cite{Beizer1990}.  \emph{Hazard
  analysis} as an activity in any safety engineering life cycle deals
with \emph{causal reasoning}, \ies establishing causal relationships
between events. Causal reasoning can be \emph{inductive}, \ies from 
causal events up to their effect events, or \emph{deductive}, \ies
from effect events down to their causal events. We focus on three
widely used techniques:
\begin{itemize}
\item \emph{fault-tree analysis (FTA)}, which looks at \emph{critical
    paths}, \ies combinations of causal factors leading to an
  undesired event with high probability~\cite{Ericson2015},
\item \emph{failure-mode-effects-analysis (FMEA)}, which helps
  identify failure modes of items and their
  propagation towards hazardous system-level
  effects~\cite{Ericson2015}, and
\item \emph{system-theoretic process analysis (STPA)}, one of the most
  recently developed techniques, which applies control structure
  models and control action guide words to determine causal
  factors~\cite{Leveson2012}.
\end{itemize}
For a detailed description of FTA and FMEA techniques see \cite{Ericson2015} and for STPA
 see \cite{Leveson2012}. In \Cref{sec:patterns:intermediate}, we provide
descriptions for the non-expert reader to gain a further
understanding.

\subsection{Motivation and Challenges}
\label{sec:motivation}
Fenelon et al.~\cite{Fenelon1994} discuss a method %
where they intertwine software HA with design increments
as countermeasures for identified hazards.  Hawkins et
al.~\cite{Hawkins2015} elaborate a tool chain for
constructing safety cases from modeled design increments.

Safety arguments are required to be valid with high
confidence~\cite{Wassyng2010,Weinstock2013}. The achievement of this
depends on the amount of details an argument should
have~\cite{2015_graydon_epistemology}. Staying too abstract is among
the many obstacles to the achievement of high confidence. We
point at several problems, \egs solutions based on FTA can refer to
large packages~\cite[pp.~319f,136]{Kelly1998}, %
argumentation strategies can encompass fairly complex reasoning
steps~\cite[pp.~17ff]{Alexander2007}\cite[p.~102]{Dardar2014}, context
and assumption statements can lack fine-granular
traceability~\cite[p.~111]{Dardar2014}.
These examples show that it is difficult to build up a comprehensible
and traceable safety argument from the HA results and
the corresponding reactions.  These problems can
\begin{itemize}
\item significantly reduce the confidence we are able to associate
  with such an argument and, consequently,
\item hinder systematic reuse of proven arguments.
\end{itemize}
Related issues were recognized by, \egs Yuan and
Xu~\cite{Yuan2010}.

\subsection{Contributions}
\label{sec:contrib}
In this paper, we seek to reduce the mentioned issues and improve
safety case methodology.  Argument patterns based on evidence from FTA
and FMEA have been discussed in~\cite{GSN2011, Alexander2007}.
Confidence arguments have to provide insight on how this evidence was
generated~\cite{2015_graydon_epistemology}.
In addition to these patterns, we discuss a pattern that \emph{zooms
  into the evidence} mentioned by the existing approaches by using
causal reasoning evidence and the evidence from developing mitigations
of identified hazards.
Our pattern helps at collecting more evidence on how the analysis is
performed.  Instead of creating an argument from error-prone
experience, we propose a \emph{systematic way} of guiding the argument
from well understood steps performed in HA to \emph{increase the
  trustworthiness} of the evidence. Hence, the structure of our
pattern resembles the causal reasoning of HA.
Moreover, the pattern \emph{adds value to HA results} by integrating
these results with the additional steps required in the construction
of confident safety cases.
We provide \emph{reusable} argument modules for different assurance
concerns, particularly, arguing for a system redesign based
on the identified hazards.

\paragraph*{Outline}
\Cref{sec:patterns} describes a pattern using goal structures to
represent safety arguments from FTA, FMEA, and STPA to substantiate
safety cases.  In \Cref{sec:runningexample}, we provide an example of
a train door control system as a basis for applying this pattern.
\Cref{sec:disc} discusses several aspects of the pattern. We close our
paper with an analysis of related work (\Cref{sec:relwork}) and
conclusions in \Cref{sec:conc}.

\section{A Modular Argument Pattern}
\label{sec:patterns}

\begin{figure}
  \centering
  \subfloat[Module overview]{
    \includegraphics[width=.48\columnwidth]{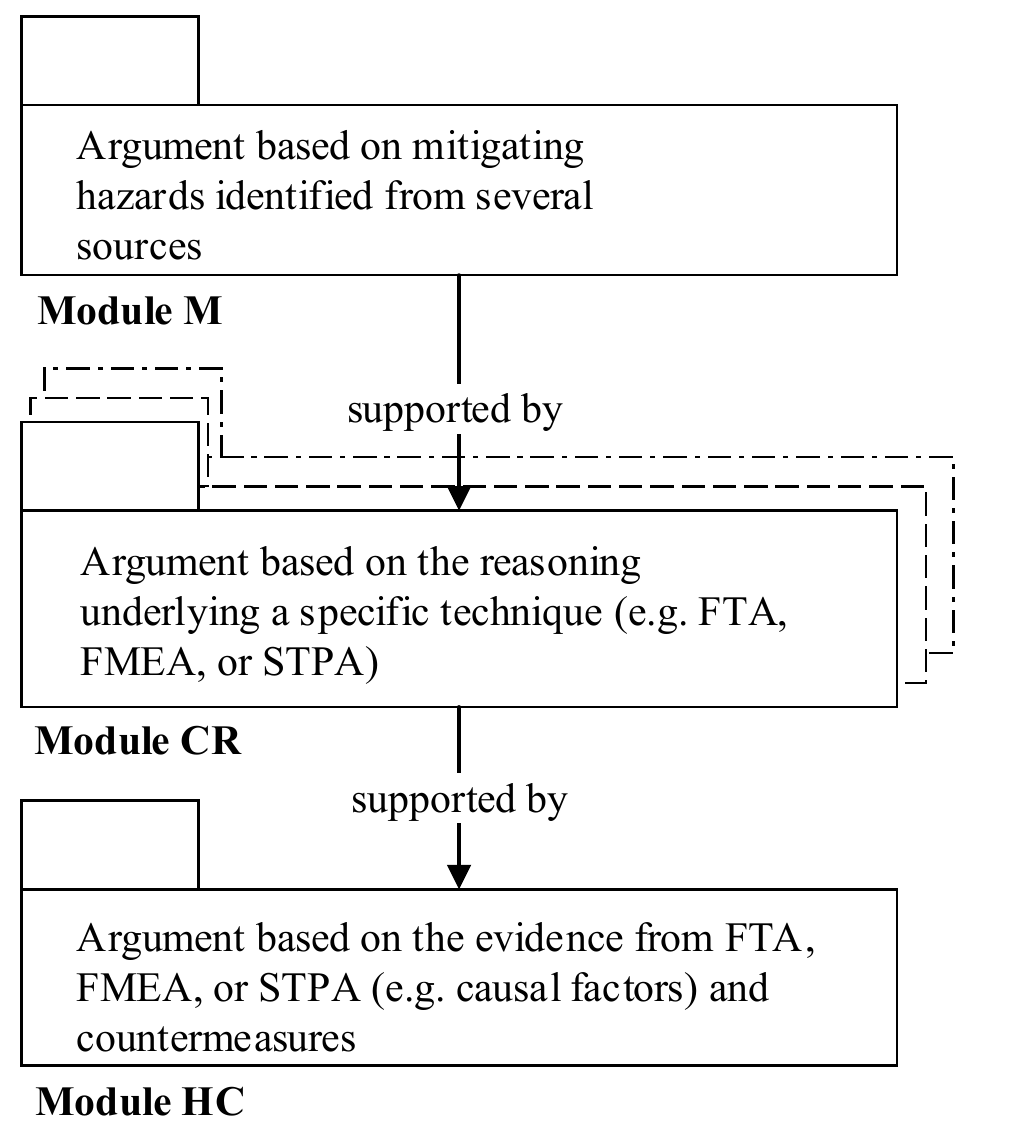}
    \label{fig:patternoverview}
  }
  \subfloat[Argument pattern for module M]{
    \includegraphics[width=.48\columnwidth]{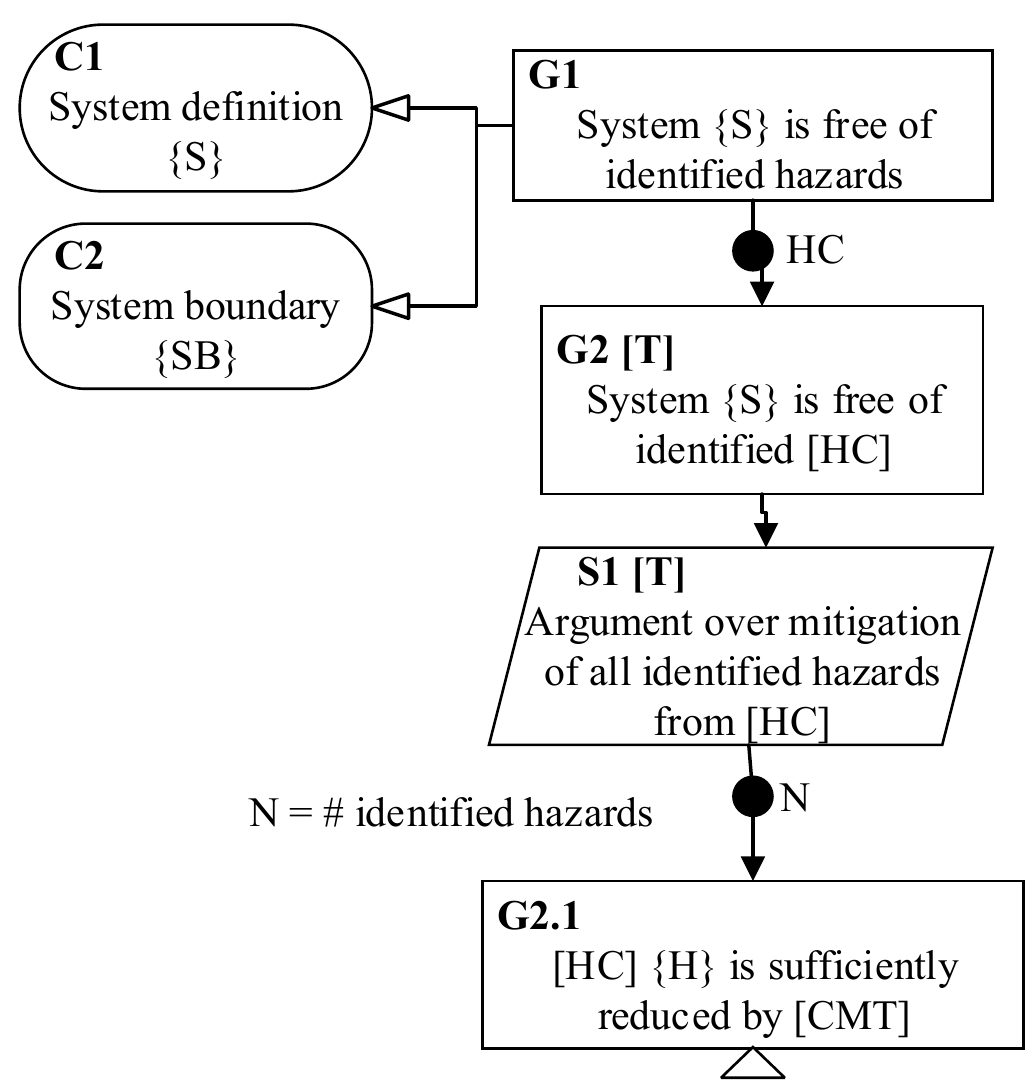}
    \label{fig:pattern:m}
  }
  \caption{Module overview and main module M}
  \vspace{-.5cm}
\end{figure}

\Cref{fig:patternoverview} describes an argument pattern in three
modules: A \textbf{Module M} describing the part of a safety argument
based on mitigating hazards from various categories, a \textbf{Module CR}
depicting the part of an argument underlying a specific HA technique,
and a \textbf{Module HC} to form detailed arguments from the reactions
after applying a specific technique.

\subsection{Concepts, Notation, and Assumptions}
\label{sec:preliminaries}

\paragraph*{Concepts}
\label{sec:concepts}

All modules are built around the concepts of
\begin{itemize}
\item \emph{hazard}, \egs failure mode or effect in
  FMEA; minimum cutset, critical path, or top-level event in FTA;
  causal factor or unsafe control action in STPA; and
\item \emph{countermeasure}, \egs corrective action in
  FMEA or safety constraint in STPA (see \Cref{sec:patterns:top}
  for further concepts).
\end{itemize}
Countermeasures can range from
\begin{inparaenum}
\item a \textbf{specification} document containing \emph{functional or
    quality requirements} and \emph{design constraints} (to be
  implemented by hardware, software, the human machine interface or
  the operator), over
\item corresponding \textbf{design changes} based on system models or
  implementation artifacts such as the application of
  \emph{dependability design patterns}, to
\item \textbf{process requirements} specifying work steps such as
  \emph{design review, verification, test}, or \emph{maintenance} to
  be conducted at certain points in the system life cycle.
\end{inparaenum}

\begin{figure}[t]
  \centering
  \includegraphics[width=\columnwidth]{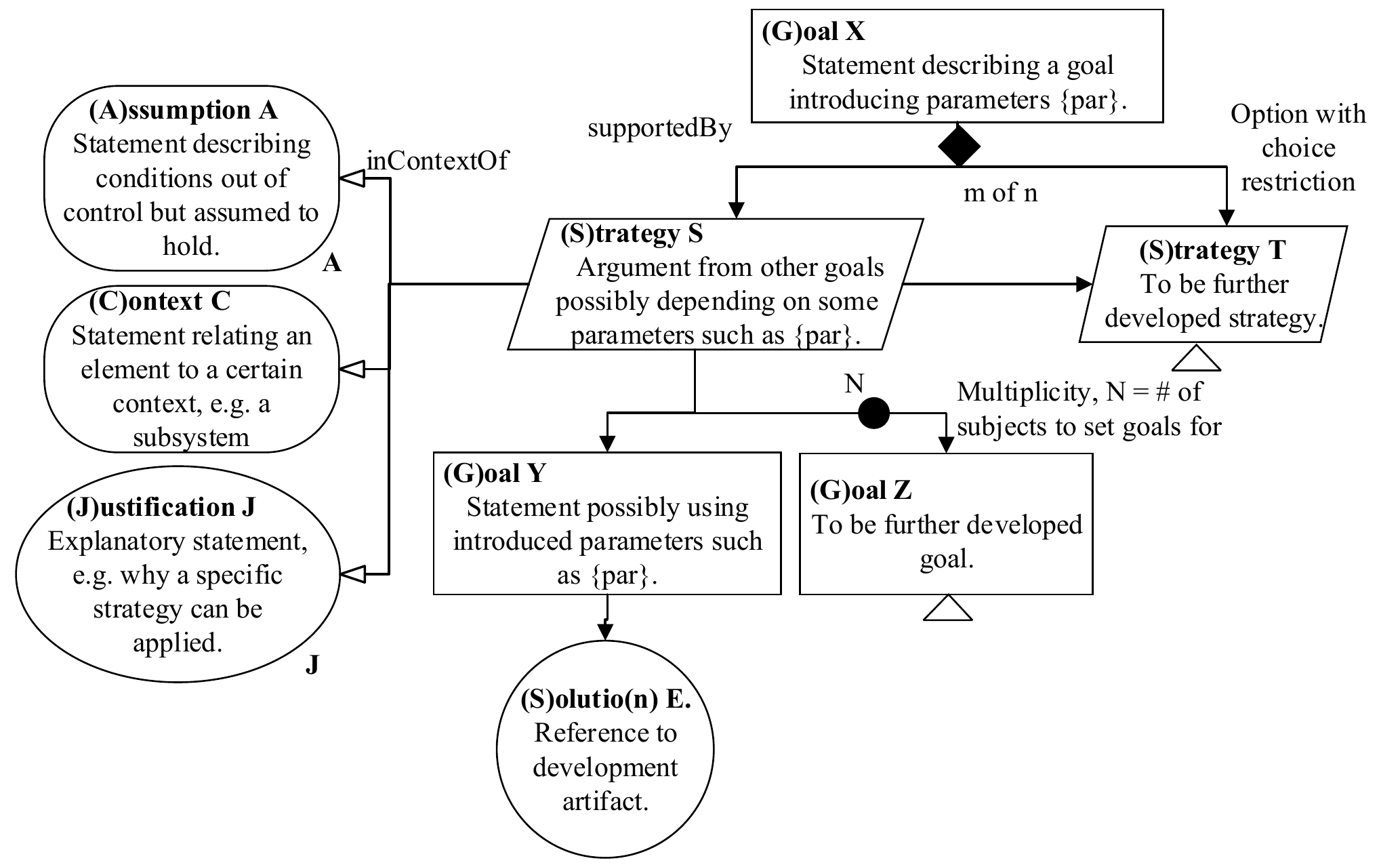}
  \caption{GSN legend. Nodes contain element descriptions.}
  \label{fig:gsn:notation}
  \vspace{-.5cm}
\end{figure}

\paragraph*{Notation}
The notation, we use in
\Crefrange{fig:pattern:m}{fig:pattern:hc:fmea:tdcs}, complies with the
Goal Structuring Notation (GSN, \cite{GSN2011}) and is described in
\Cref{fig:gsn:notation}.  We distinguish between parameters for
pattern refinement (indicated by square brackets ``[]'') and for
pattern instantiation (indicated by curly brackets ``\{\}'').
GSN is one way of visualizing argumentation in safety cases. However,
the following approach should, in principle, work with any means used
to visualize assurance argumentation.

\paragraph*{Assumptions}
Similar to Hawkins et al.~\cite{Hawkins2015}, we assume model-based
development to be a basis for constructing the parts of the argument
referring to system design increments.  Hence, these increments refer
to a system model describing design-related safety measures to be
implemented. The implementation then needs to be verified against the
model and the requirements to complete the argument.

\subsection{Module M: Arguing from Hazard Mitigation}
\label{sec:patterns:top}

\begin{figure}[t]
  \centering
  \subfloat[Strategy 1 for FTA and STPA]{
    \includegraphics[width=.48\columnwidth]{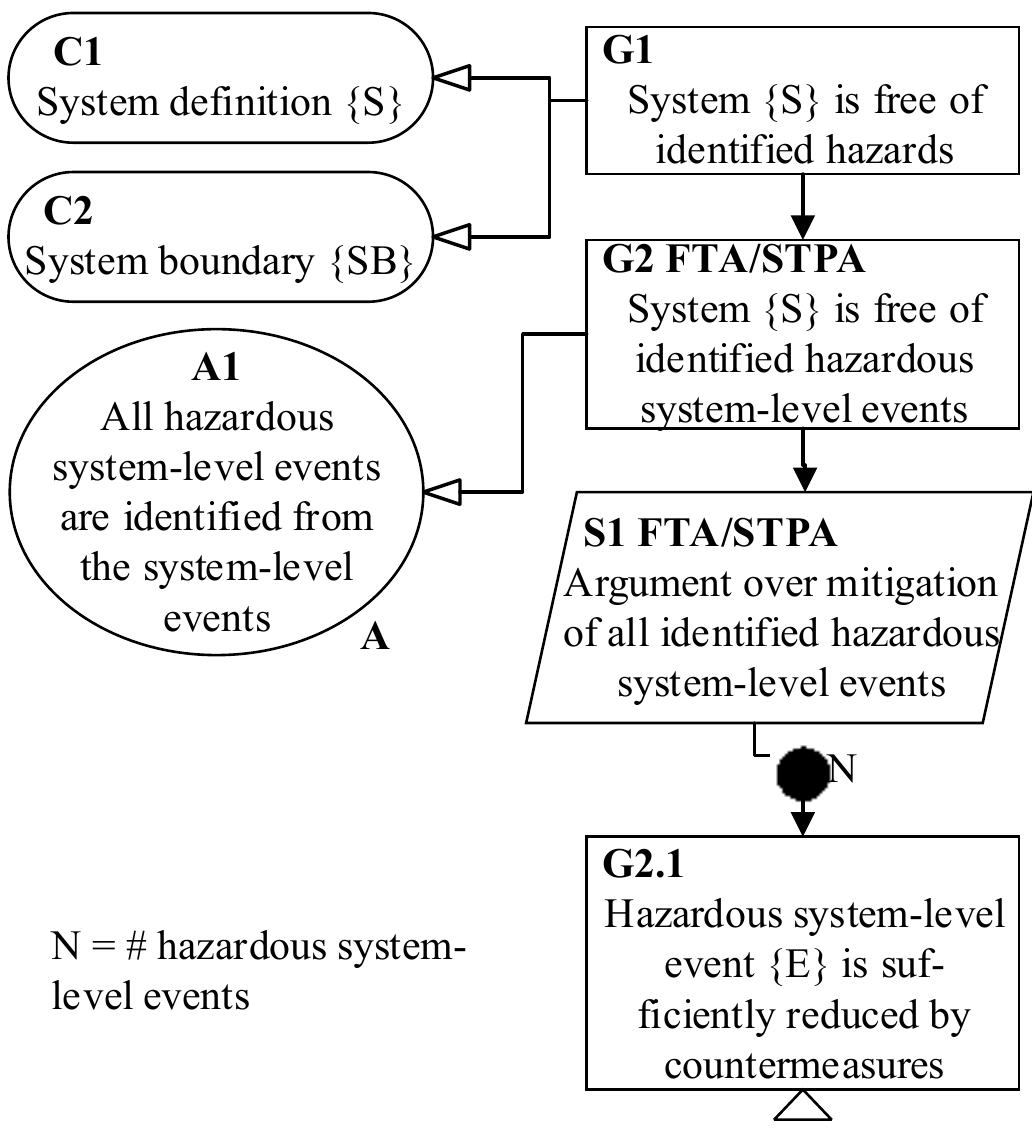}
    \label{fig:pattern:m:fta}
  }
  \subfloat[Strategy 1 for FMEA]{
    \includegraphics[width=.48\columnwidth]{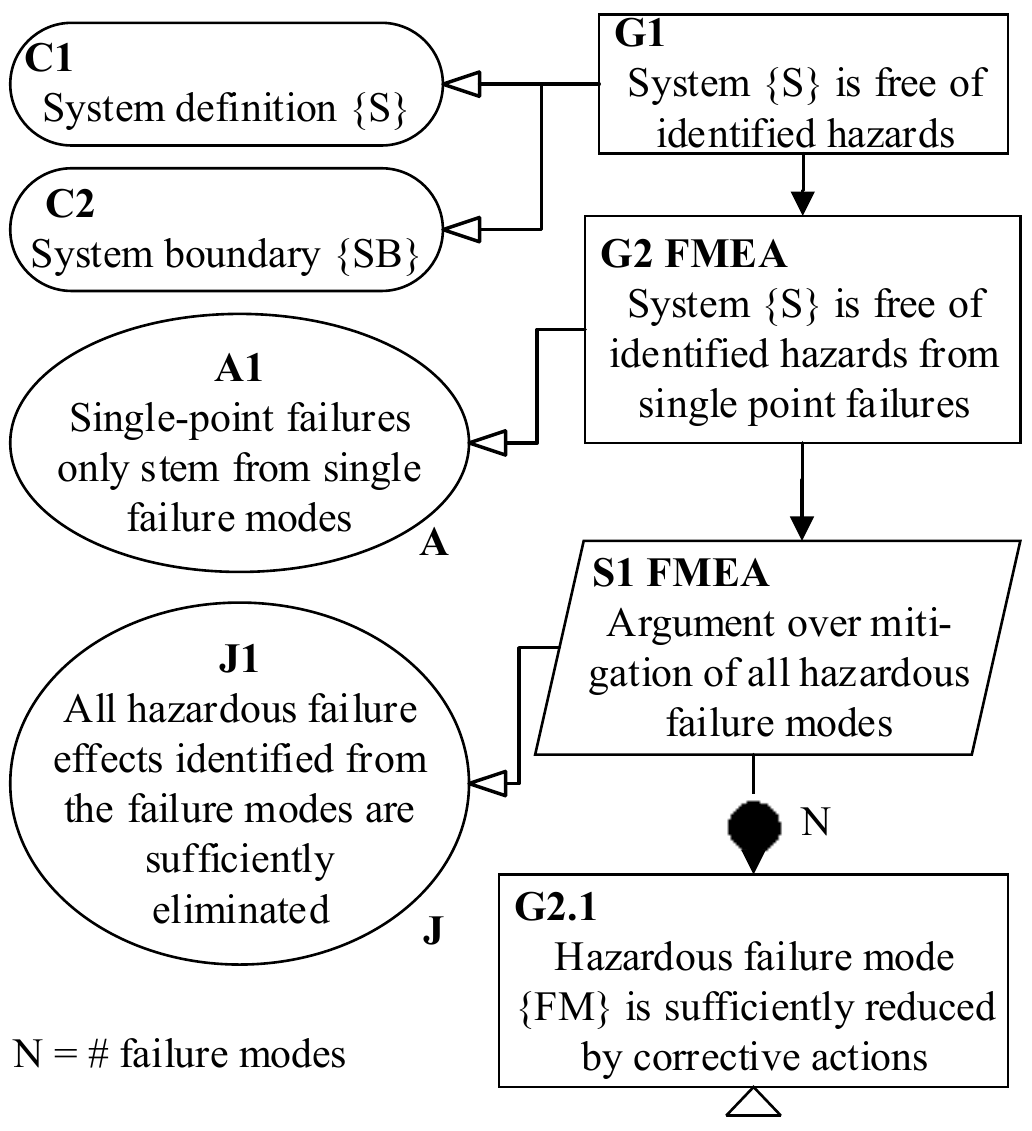}
    \label{fig:pattern:m:fmea}
  }
  \caption{Refinements of \Cref{fig:pattern:m} for FTA, STPA, and FMEA}
\end{figure}

Module $M$ describes an argument pattern which aims at coverage of two
main categories of hazards we focus\footnote{Motivated from
  \Cref{sec:safetycases}, these two categories cover a large subclass
  of identifiable and important hazards.} on, \ies \emph{hazardous
  single-point\footnote{For sake of simplicity, we do not discuss
    common cause failures which, however, can be covered by specific
    variants of FMEA and FTA.} failures} (objective of FMEA) and
\emph{hazardous system-level events} (objective of FTA and STPA). The
goal structure in \Cref{fig:pattern:m} therefor contains a
multiplicity over the parameter \texttt{HC} (hazard
category). Furthermore, the parameter \texttt{CMT} ranges over the
terms \emph{countermeasure (universal), design revision (universal),
  safety constraint (STPA)}, \emph{corrective action (FMEA)}, and
\emph{process-based measure (FMEA)}.
\Cref{fig:pattern:m:fta,fig:pattern:m:fmea} refine module $M$ via
\texttt{HC}.

\subsection{Module CR: Arguing from Causal Reasoning}
\label{sec:patterns:intermediate}

Here, we provide three argument patterns incorporating the type of
causal reasoning underlying FTA, STPA, and FMEA. 

\subsubsection{Commonalities among the Techniques}
\label{sec:comm-among-techn}
By the ``1-out-of-2'' choice in
\Cref{fig:pattern:cr:fta,fig:pattern:cr:fmea,fig:pattern:cr:stpa}, the
pattern allows to complement \emph{design revisions} with
\emph{process-based measures}, \egs unit tests, material quality
checks, reviews.  This distinction allows us to construct an argument
along these two commonly used lines of argumentation (see, \egs
\cite{Leveson2011}).

\subsubsection{Argument based on FTA}
\label{sec:pattern:fta}

\begin{figure}[t]
  \centering
  \includegraphics[width=\columnwidth]{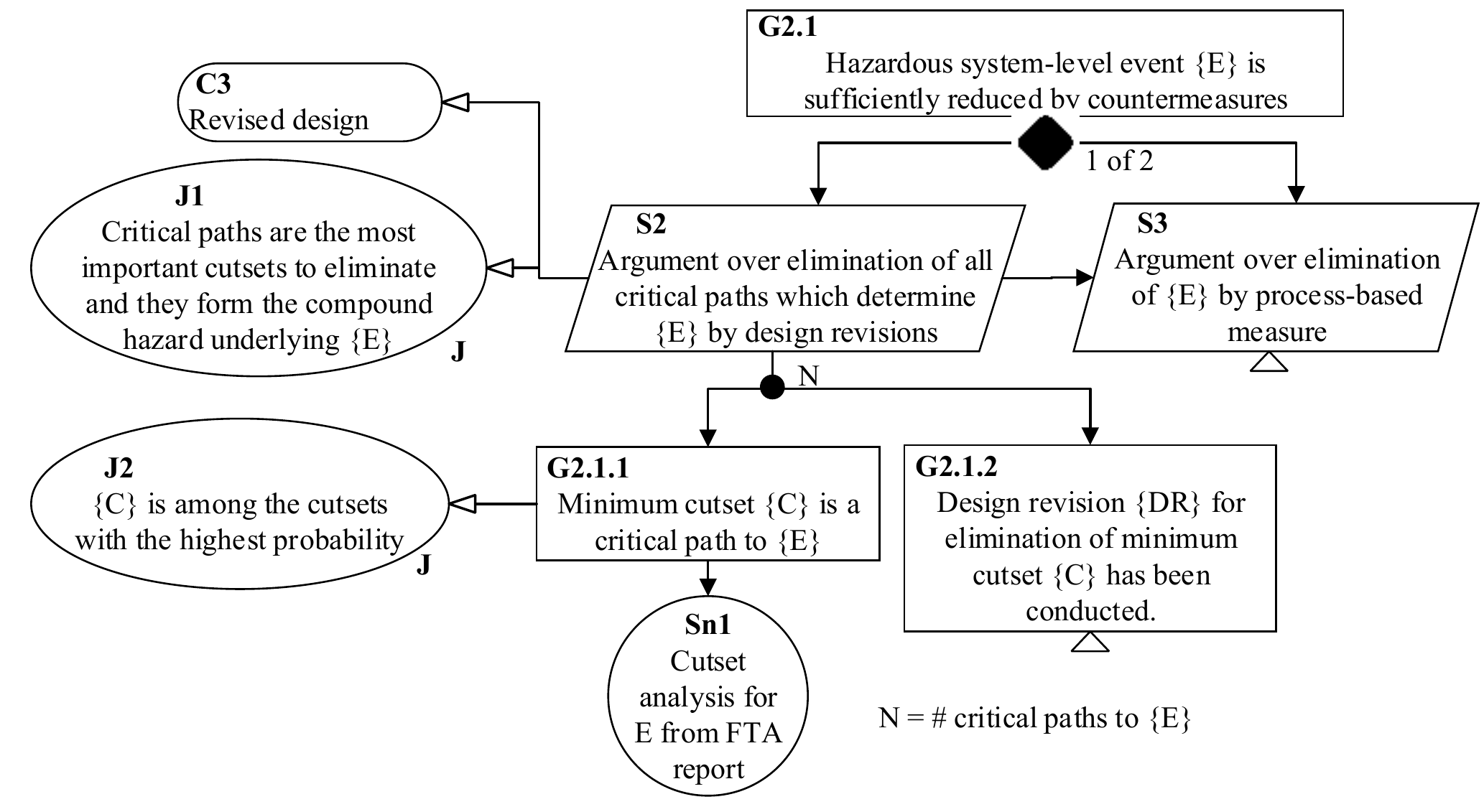}
  \caption{Module CR argument pattern for FTA}
  \label{fig:pattern:cr:fta}
  \vspace{-.7cm}
\end{figure}

The main part of the argument in \Cref{fig:pattern:cr:fta} contains
the \emph{elimination of all critical paths} (Strategy 2), \ies the
minimum cutsets $C$ with comparably high probabilities, leading to an
undesired system-level event $E$.  Strategy 2 can be used to construct
arguments over the (i) \emph{elimination of critical paths}, or (ii)
the \emph{reduction of failure rates of these critical paths} below
an acceptable maximum. Both approaches, (i) and (ii), are associated with
a \emph{design revision} $DR$. The multiplicity helps
to argue over arbitrarily many relevant pairs $(C,DR)$.  By \emph{C2}
in \Cref{fig:pattern:m:fta}, we consider system-level events $E$ as
top-level events of an FTA.

\paragraph*{Notes on Construction}
As opposed to the \emph{fault-tree evidence} pattern described in
\cite[pp.~186f]{Kelly1998} and following FTA deductive causal
reasoning, we consider this pattern being
used to construct an argument top-down from Goal 2.1.

\subsubsection{Argument based on FMEA}
\label{sec:pattern:fmea}

\begin{figure}[t]
  \centering
  \includegraphics[width=\columnwidth]{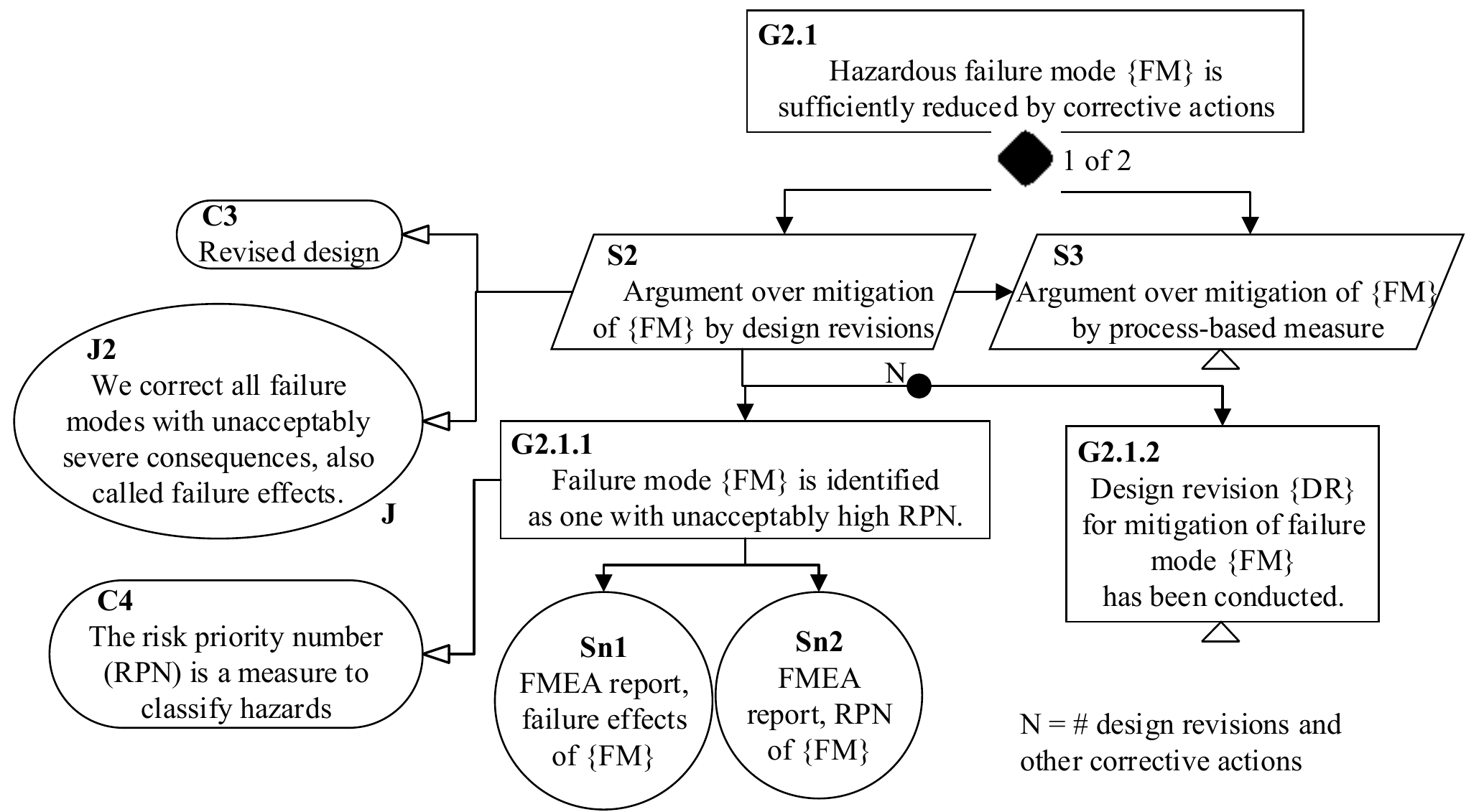}
  \caption{Module CR argument pattern for FMEA}
  \label{fig:pattern:cr:fmea}
\end{figure}

The main part of the argument shown in \Cref{fig:pattern:cr:fmea}
pertains \emph{Strategy 2} which argues over a correctly identified
failure mode $FM$ with high risk priority number (RPN) and a
corresponding design revision $DR$ taken to mitigate $FM$.  The
multiplicity allows $FM$ to be mitigated by arbitrarily many recorded
design revisions.

\paragraph*{Notes on Construction}
We consider this pattern being used to construct an
argument top-down from Goal 2.1. This $CR$ pattern, particularly
\emph{Strategy 2}, inverts the FMEA inductive causal reasoning in the
sense that the elimination of hazardous failure effects (Goal 2) from
failure modes is established by building up the goal structure in a
top-down manner, \ies descending from Goal 2 towards Goal 2.1.1.

\subsubsection{Argument based on STPA}
\label{sec:pattern:stpa}

To reuse Module $M$ from $FTA$, we use the concepts hazard $H$ and
accident $A$ to form the hazardous system-level event $E$.  In the
context of STPA, \emph{Goal 2.1} can be interpreted as a reduction of
the probabilities of the identified accidents $A$ by mitigation of
their corresponding hazards $H$.

\begin{figure}[t]
  \centering
  \includegraphics[width=\columnwidth]{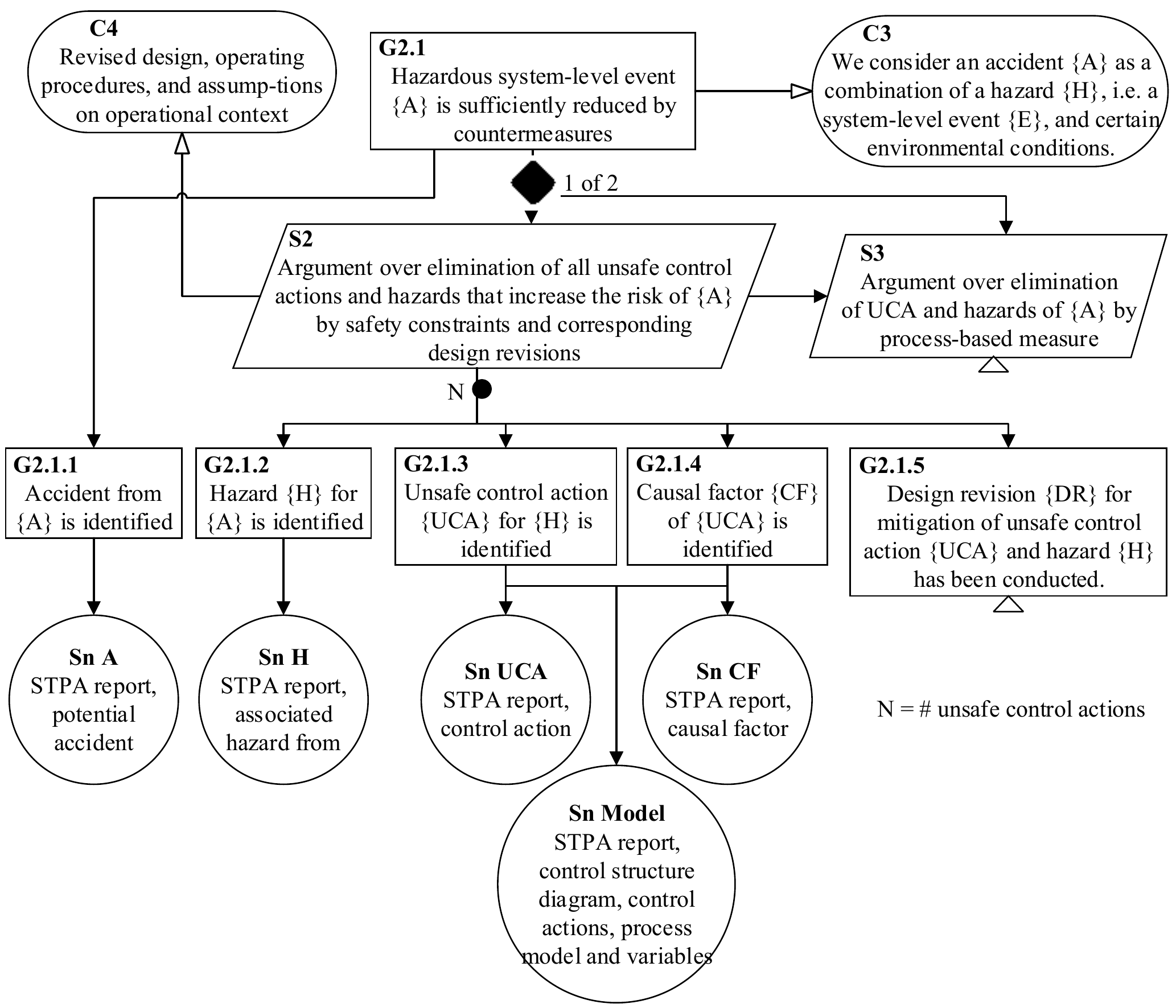}
  \caption{Module CR argument pattern for STPA}
  \label{fig:pattern:cr:stpa}
  \vspace{-.6cm}
\end{figure}

The CR pattern for STPA (\Cref{fig:pattern:cr:stpa}) works
similar to the one for FTA (\Cref{fig:pattern:cr:fta}). However, it is
more elaborate in terms of the concepts used to decompose the chain of
evidence in causal reasoning and the goals. Based on an identified
accident $A$, \emph{Strategy 2} argues over the elimination of
\emph{all unsafe control actions $UCA$ and corresponding hazards $H$}
(Goals 2.1.2 to 2.1.4) by design measures (Goal 2.1.5).

\paragraph*{Notes on Construction}
Similar to the FTA deductive causal reasoning, we consider this
pattern mainly being applicable in top-down argumentation from Goal
2.1. However, the \textbf{Goals 2.1.1 to 2.1.4} might be elaborated
\begin{enumerate}
\item from left to right, \egs if we want to follow STPA starting with
  accident analysis, or 
\item from right to left, \egs if we want to study the causal factors
  resulting from a former safety case or a former accident
  investigation and follow the course of events.
\end{enumerate}

\subsection{Module HC: Arguing from Hazard Countermeasures}
\label{sec:patterns:bottom}

The goal structures of this module argue over evidence from results of
the applied (i) HA techniques, \ies hazards, and (ii)
assurance techniques, \ies countermeasures (see \Cref{sec:concepts}).
Now, we describe a generic argument pattern for module HC
followed by three refinements for FTA, FMEA, and STPA.

\subsubsection{Generic Argument}
\label{sec:countermeasures-generic}

\begin{figure}[t]
  \centering
  \includegraphics[width=\columnwidth]{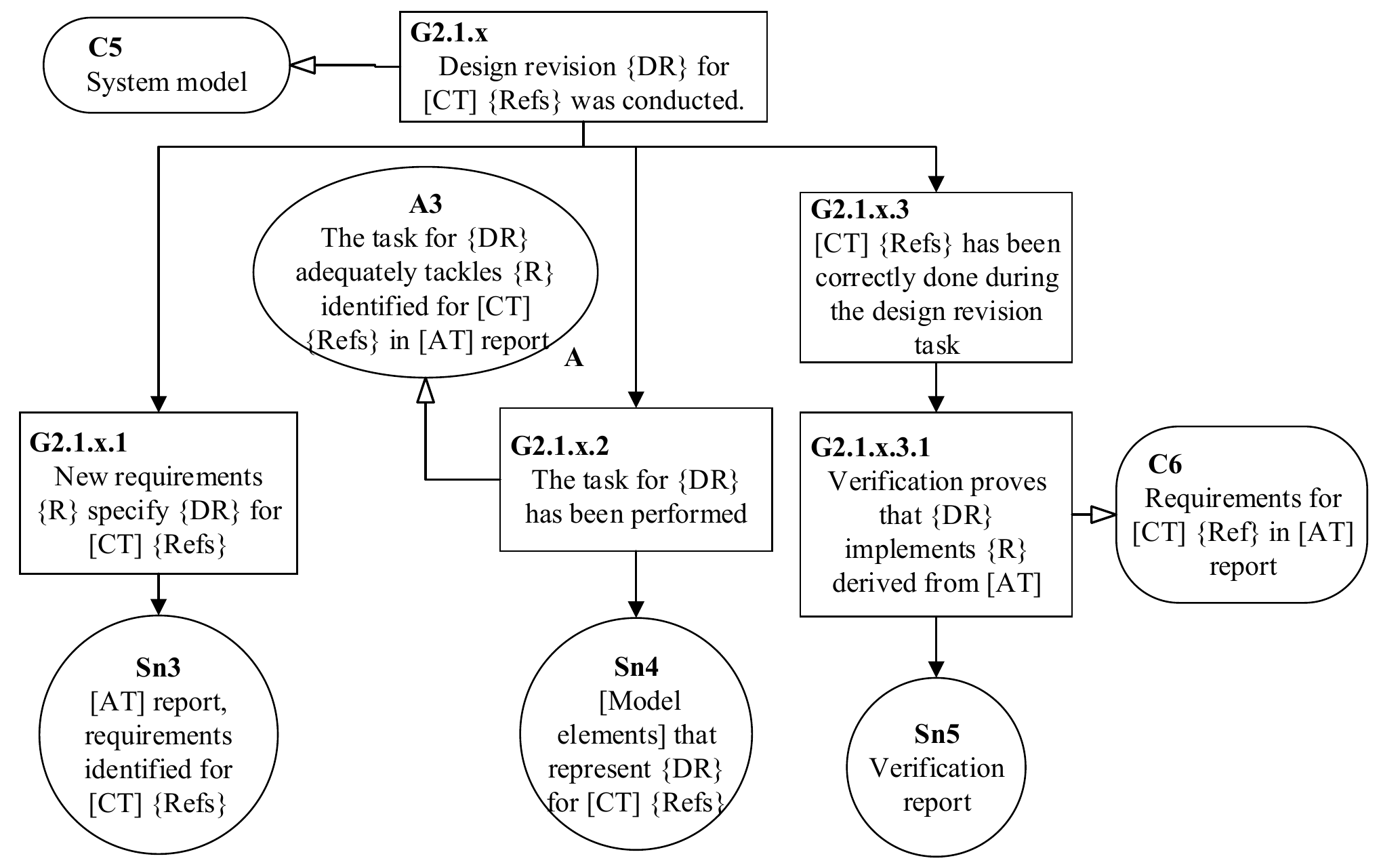}
  \caption{Generic argument pattern for module HC}
  \label{fig:pattern:hc}
  \vspace{-.3cm}
\end{figure}

In \Cref{fig:pattern:hc}, the \textbf{Goal 2.1.x} mentions the
parameter $CT$ which denotes the \emph{type of cause to be mitigated,
  treated, or eliminated} together with a specifier $\mathit{Refs}$ as
explained below. Furthermore, the parameter $AT$ refers to the
analysis technique from which the corresponding evidence (\ies
solutions) can be collected.  The requirements $R$ identified for the
design revision $DR$ according to Goal 2.1.x.1 are implemented by Goal
2.1.x.2 and verified after Goal 2.1.x.3.

\subsubsection{Argument based on FTA}
\label{sec:countermeasures-fta}

\begin{figure}[t]
  \centering
  \includegraphics[width=\columnwidth]{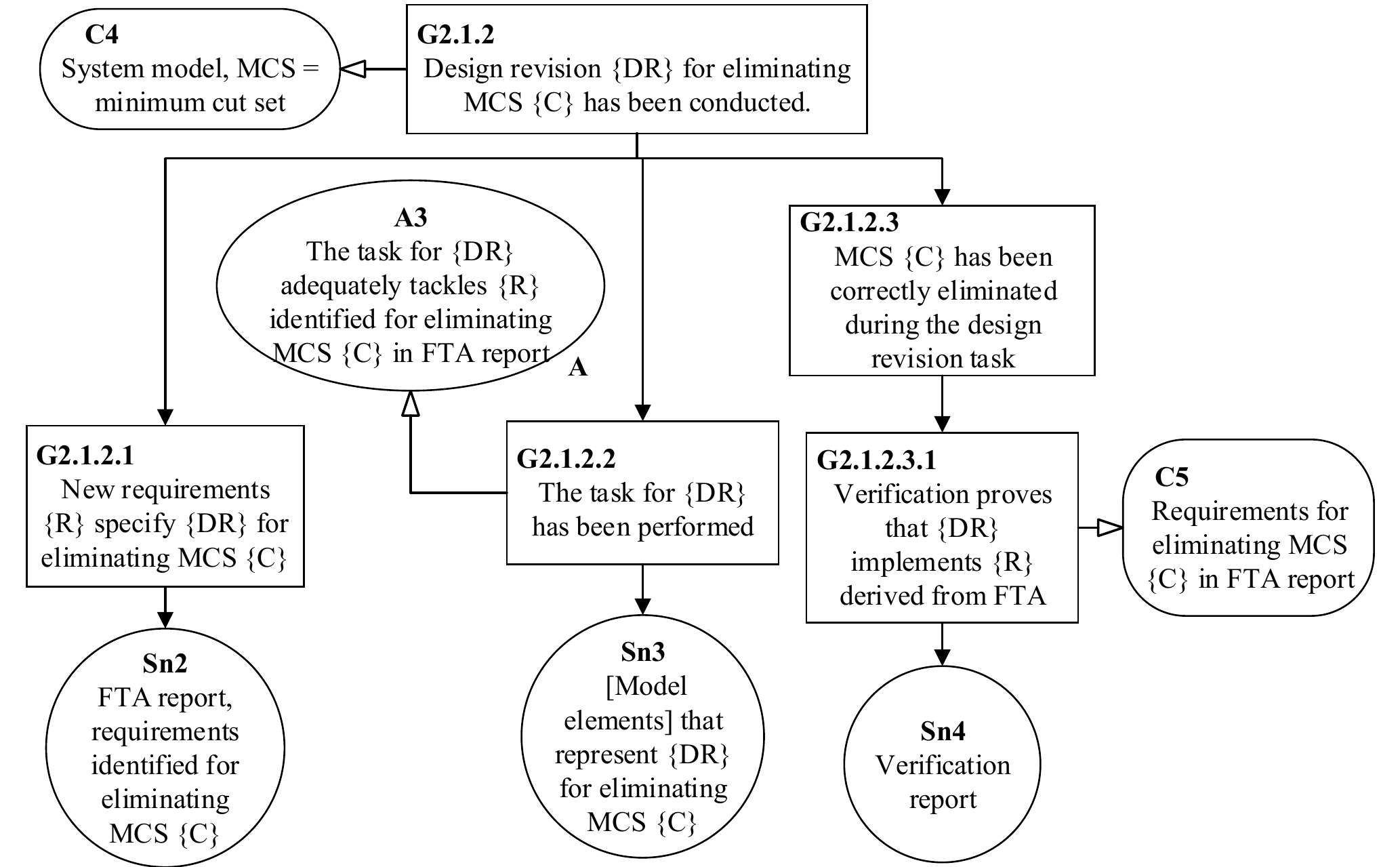}
  \caption{Refinement of \Cref{fig:pattern:hc} for FTA}
  \label{fig:pattern:hc:fta}
  \vspace{-.5cm}
\end{figure}

\Cref{fig:pattern:hc:fta} refines the argument of
\Cref{fig:pattern:hc}: The parameter $CT$ is substituted by
``eliminating minimum cutset (MCS)'' and $\mathit{Refs}$ by a
parameter $\{C\}$ to specify a minimum cutset. $AT$ is set to FTA.

\subsubsection{Argument based on FMEA}
\label{sec:countermeasures-fmea}

\begin{figure}[t]
  \centering
  \includegraphics[width=\columnwidth]{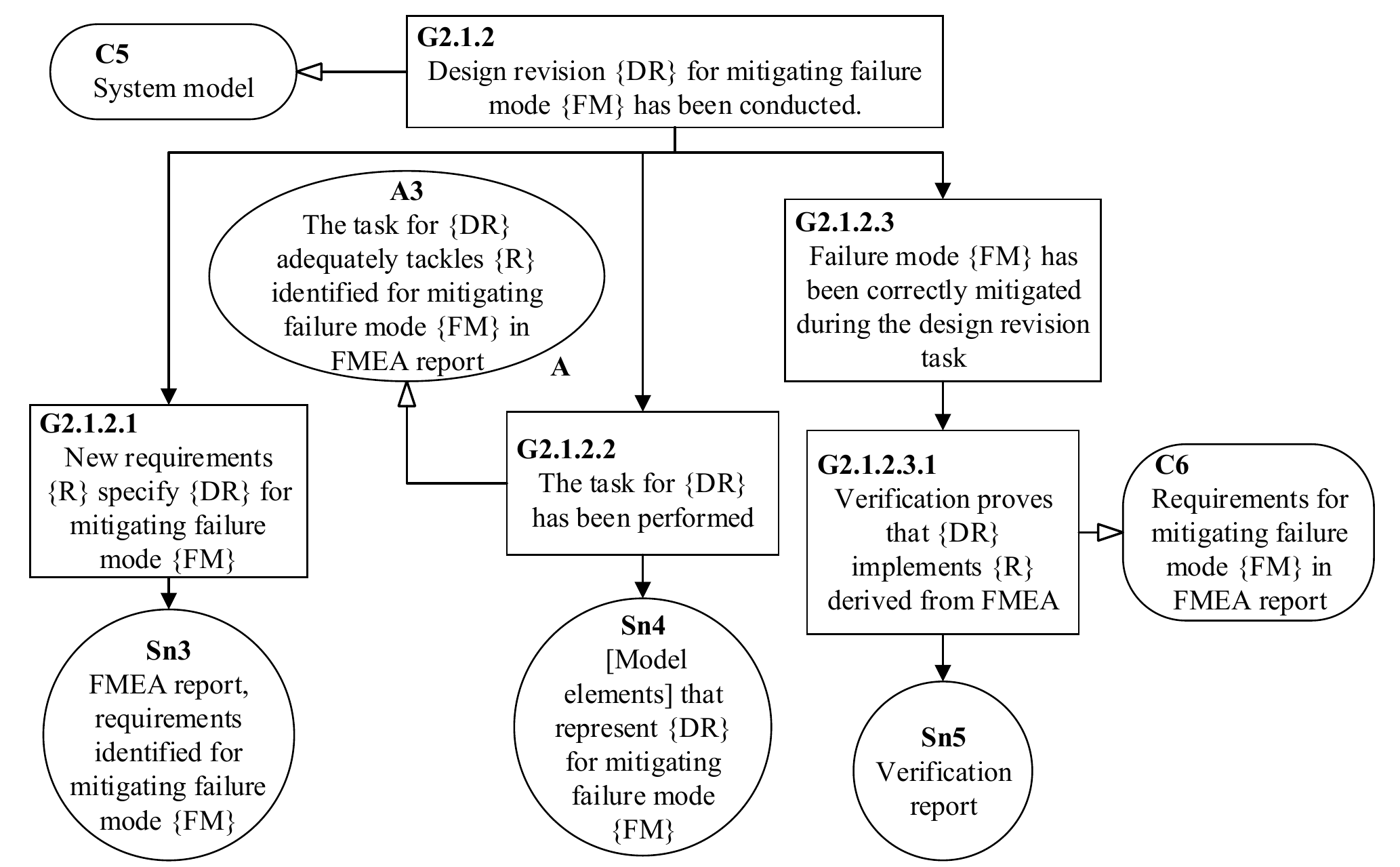}
  \caption{Refinement of \Cref{fig:pattern:hc} for FMEA}
  \label{fig:pattern:hc:fmea}
  \vspace{-.3cm}
\end{figure}

\Cref{fig:pattern:hc:fmea} refines the argument of
\Cref{fig:pattern:hc}: The parameter $CT$ is substituted by
``mitigating failure mode'' and $\mathit{Refs}$ by a parameter
$\{FM\}$ to specify a failure mode. $AT$ is substituted by FMEA.

\subsubsection{Argument based on STPA}
\label{sec:countermeasures-stpa}

\begin{figure}[t]
  \centering
  \includegraphics[width=\columnwidth]{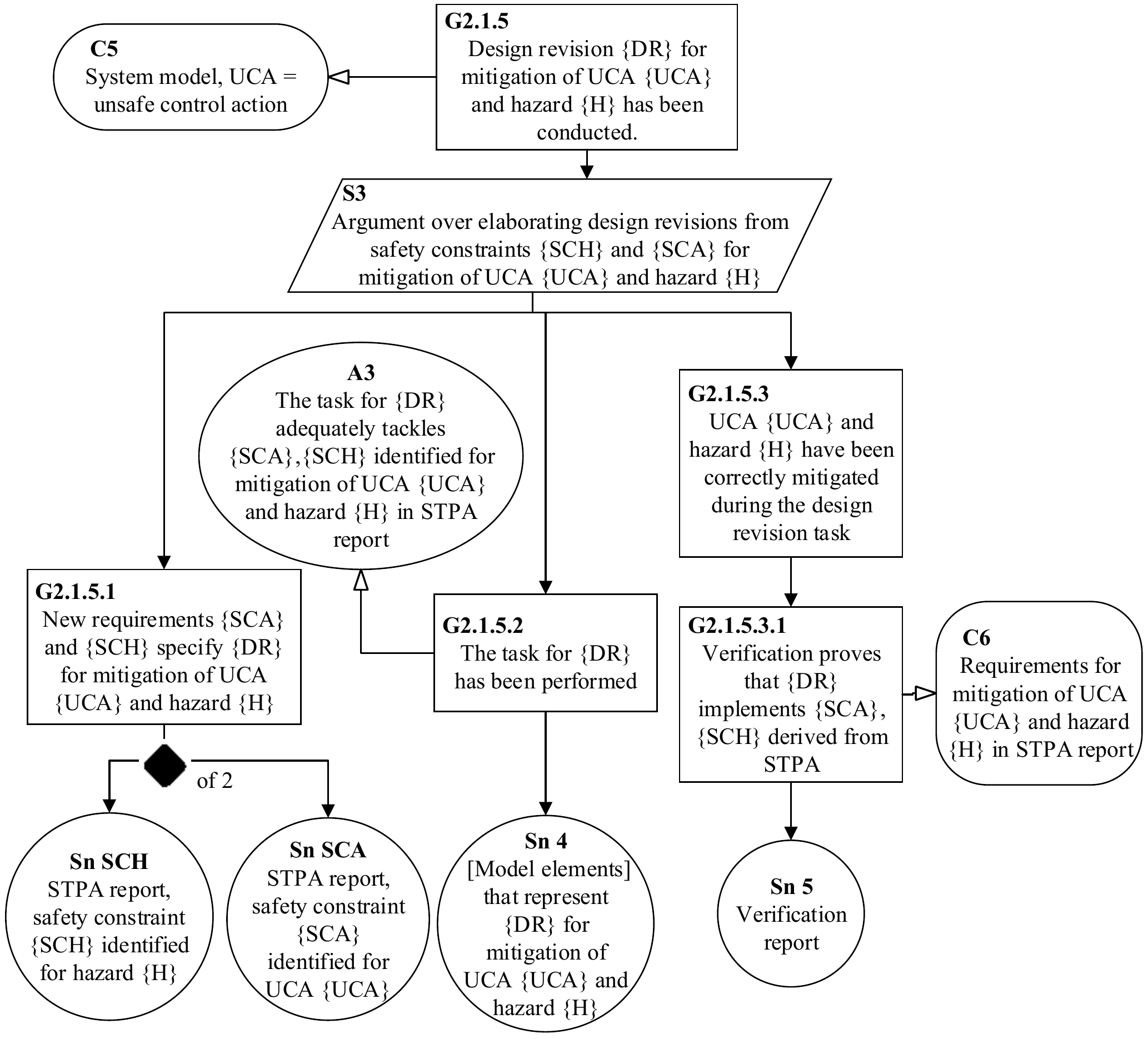}
  \caption{Refinement of \Cref{fig:pattern:hc} for STPA}
  \label{fig:pattern:hc:stpa}
  \vspace{-.5cm}
\end{figure}

In STPA, safety constraints are requirements which constrain the
control system structure and behavior such that unsafe control actions
and, in turn, hazards are mitigated at specification level.
\Cref{fig:pattern:hc:stpa} refines the argument of
\Cref{fig:pattern:hc} by setting the parameter
$CT$ to ``mitigation of unsafe control action and hazard'' and
$\mathit{Refs}$ to $\{UCA\}$ and $\{H\}$. Specific to
the work steps in STPA, we propose a ``1-out-of-2'' choice depending
on whether the safety constraints (\ies specific requirements) have
been derived from an unsafe control action $UCA$ (instantiating the
parameter $SCA$), from a hazard $H$ (instantiating the parameter
$SCH$), or from both. Finally, $AT$ is substituted by STPA.

\section{Application: Train Door Control System}
\label{sec:runningexample}

In this section, we discuss a simplified train door control system
(TDCS) as an example of a safety-critical software-based system in the
public train systems domain.
 
\subsection{Information about the System}
\label{sec:inform-about-syst}

\paragraph*{Features}
Our TDCS is responsible for operating a single door unit,
\ies \emph{opening doors} according to passengers' or train conductor's
requests, \emph{closing doors} according to train conductor's
requests, both only in appropriate situations.

\paragraph*{System-level Safety Requirements (SR)}
After train-level hazard identification, a TDCS has to
particularly fulfill train-level safety requirements such as, \egs
\begin{description}
\item[\textbf{SR1}] locking the doors closed while the train is
  moving,
\item[\textbf{SR2}] preventing the train from moving while the doors
  are not locked,
\item[\textbf{SR3}] not harming humans residing in the
  doorway,
\item[\textbf{SR4}] allowing manual opening after the train stopped in
  case of an emergency.
\end{description}

\begin{figure}[t]
  \centering
  \includegraphics[width=\columnwidth]{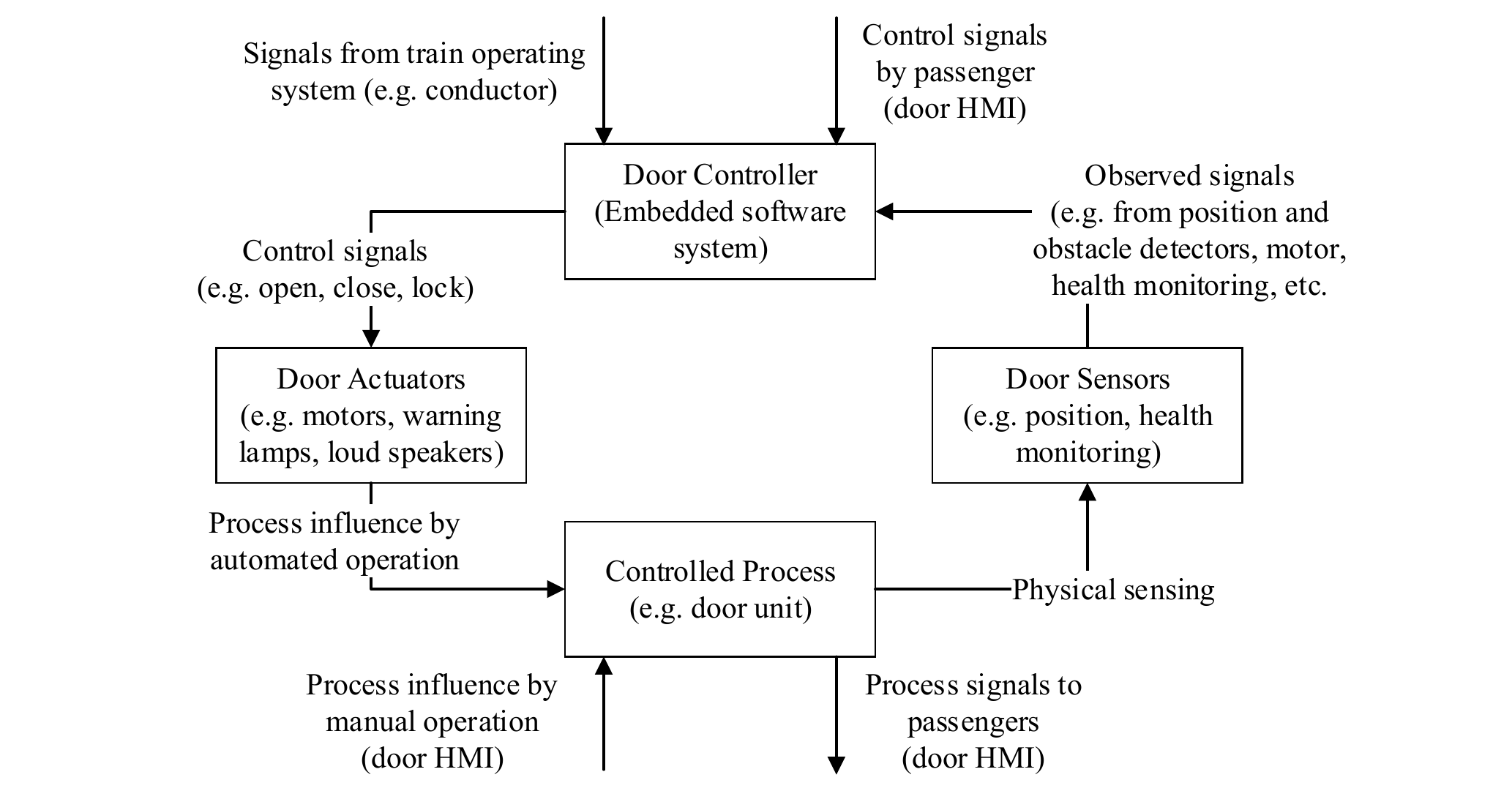}
  \caption{Control structure diagram of a simple TDCS}
  \label{fig:tdcs:structure}
  \vspace{-.5cm}
\end{figure}

\paragraph*{System Structure}
\Cref{fig:tdcs:structure} shows a simplified control loop with the
main components of the TDCS.

\subsection{Application of Module M}
\label{sec:appl:m}
We applied FTA and FMEA to our TDCS. Hence, our goal structure is
complete in the sense of capturing both causal reasoning directions,
inductive and deductive.

\begin{figure}[t]
  \centering
  \includegraphics[width=\columnwidth]{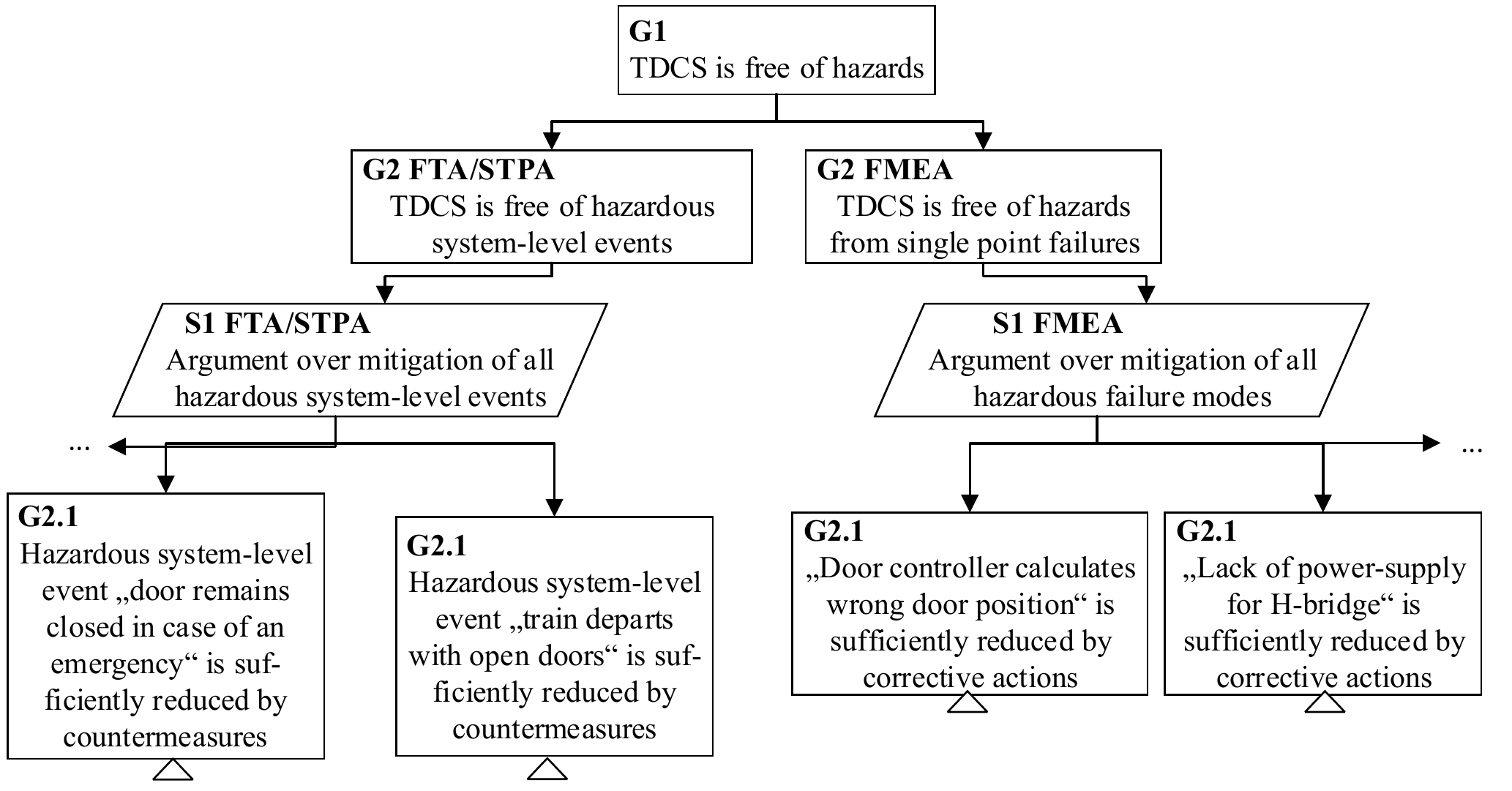}
  \caption{Application of module M to TDCS}
  \label{fig:pattern:m:tdcs}
  \vspace{-.5cm}
\end{figure}

\Cref{fig:pattern:m:tdcs} applies module M by substituting
$S$ for TDCS, using the refined patterns for FMEA and FTA for the two
hazardous system-level events ``door remains closed in case of
emergency'' (derived from SR4) and ``train departs with open doors''
(derived from SR1 and SR2), and the two failure modes ``door controller
calculates wrong door position'' and ``lack of power supply for
H-bridge.'' For sake of brevity, we left most context,
assumption, and justification elements away. Please,
consider the pattern description in \Cref{sec:patterns}.

\subsection{Application of Module CR}
\label{sec:appl-module-cr}

\begin{figure}[t]
  \centering
  \subfloat[For FTA]{
    \includegraphics[width=\columnwidth]{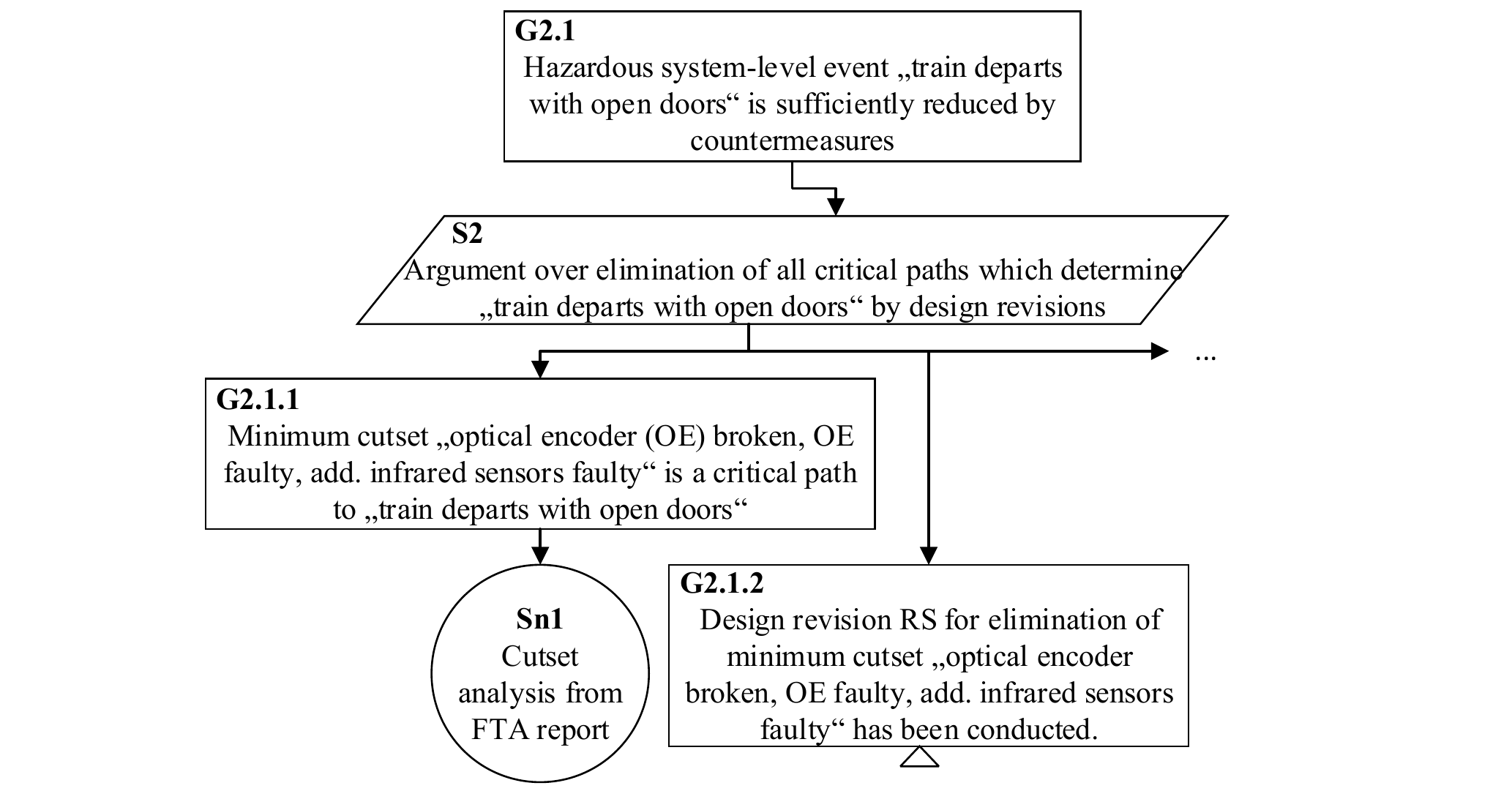}
    \label{fig:pattern:cr:fta:tdcs}
  }\\
  \subfloat[For FMEA]{
    \includegraphics[width=\columnwidth]{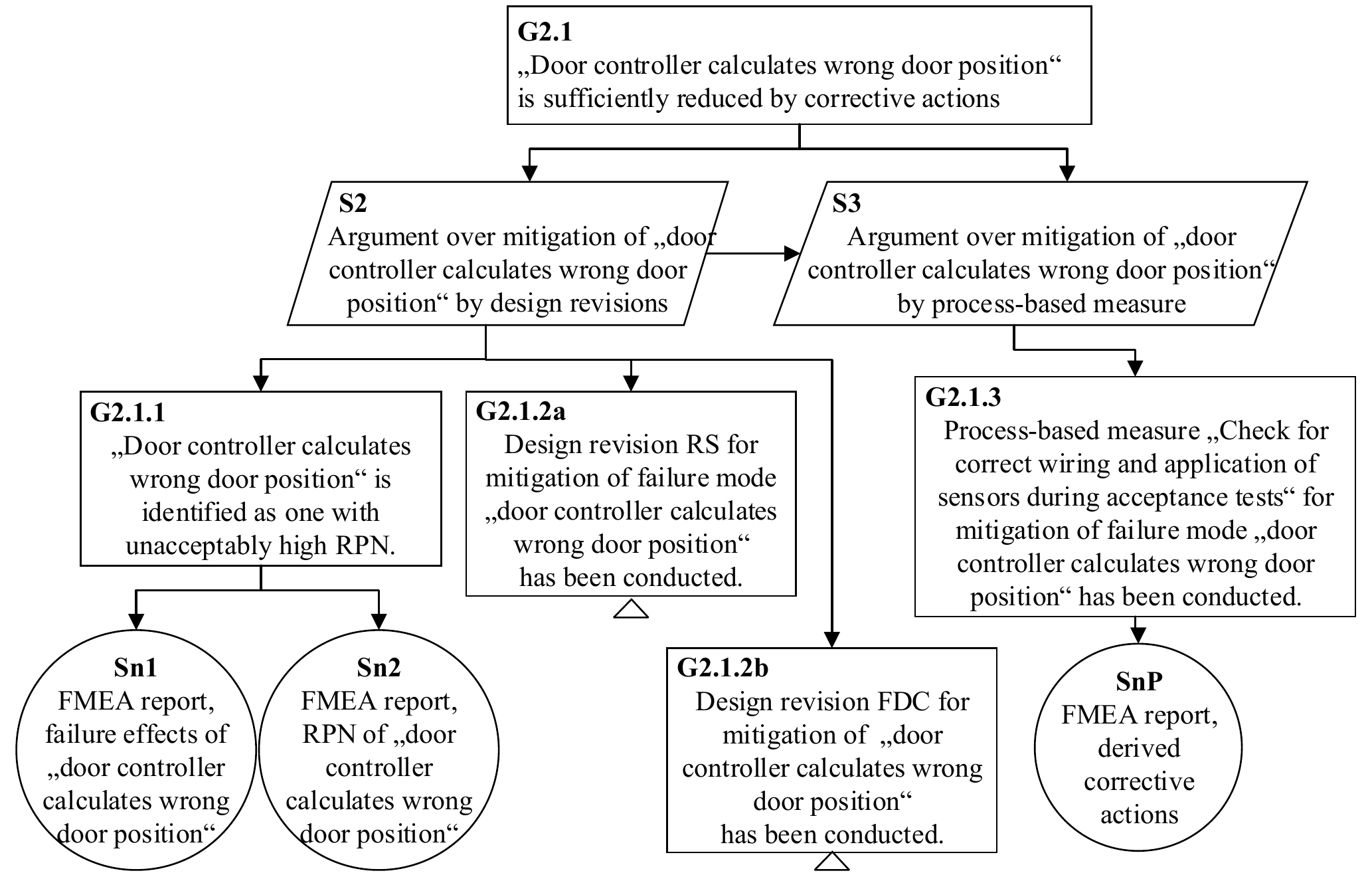}
    \label{fig:pattern:cr:fmea:tdcs}
  }
  \caption{Application of module CR to TDCS}
  \vspace{-.7cm}
\end{figure}

\paragraph*{Module CR-FTA}
\label{sec:appl:cr:fta}
In \Cref{fig:pattern:cr:fta:tdcs}, we show a breakdown of the goal
structure for the event ``train departs with open doors.'' We show the
cutset ``optical encoder broken or faulty, additional infrared sensors
faulty'' as an example. This cutset was determined a as critical path
by FTA. We only consider one critical path in this
example. \textbf{Goal 2.1.2} requires a design revision identified by
``robust sensors (RS)'' to be successfully conducted. We discuss this
in \Cref{sec:appl:hc}.

\paragraph*{Module CR-FMEA}
\label{sec:appl:cr:fmea}
\Cref{fig:pattern:cr:fmea:tdcs} indicates how FMEA enriches our
argument. For \textbf{Goal 2.1}, we pick the failure mode ``door
controller calculates wrong door position'' whose reduction is argued
by two strategies: \textbf{Strategy 2} builds on correct
identification of this failure mode (Goal 2.1.1) and on two design
revisions $RS$ and $FDC$ (Goals 2.1.2a and b). \textbf{Strategy 3}
builds on measures in the system integration process, \ies a ``check
for correct wiring and sensor application'' (Goal 2.1.3).  Again, we
only consider one failure mode.

\subsection{Application of Module HC}
\label{sec:appl:hc}

\begin{figure}[t]
  \centering
  \subfloat[For FTA]{
    \includegraphics[width=\columnwidth]{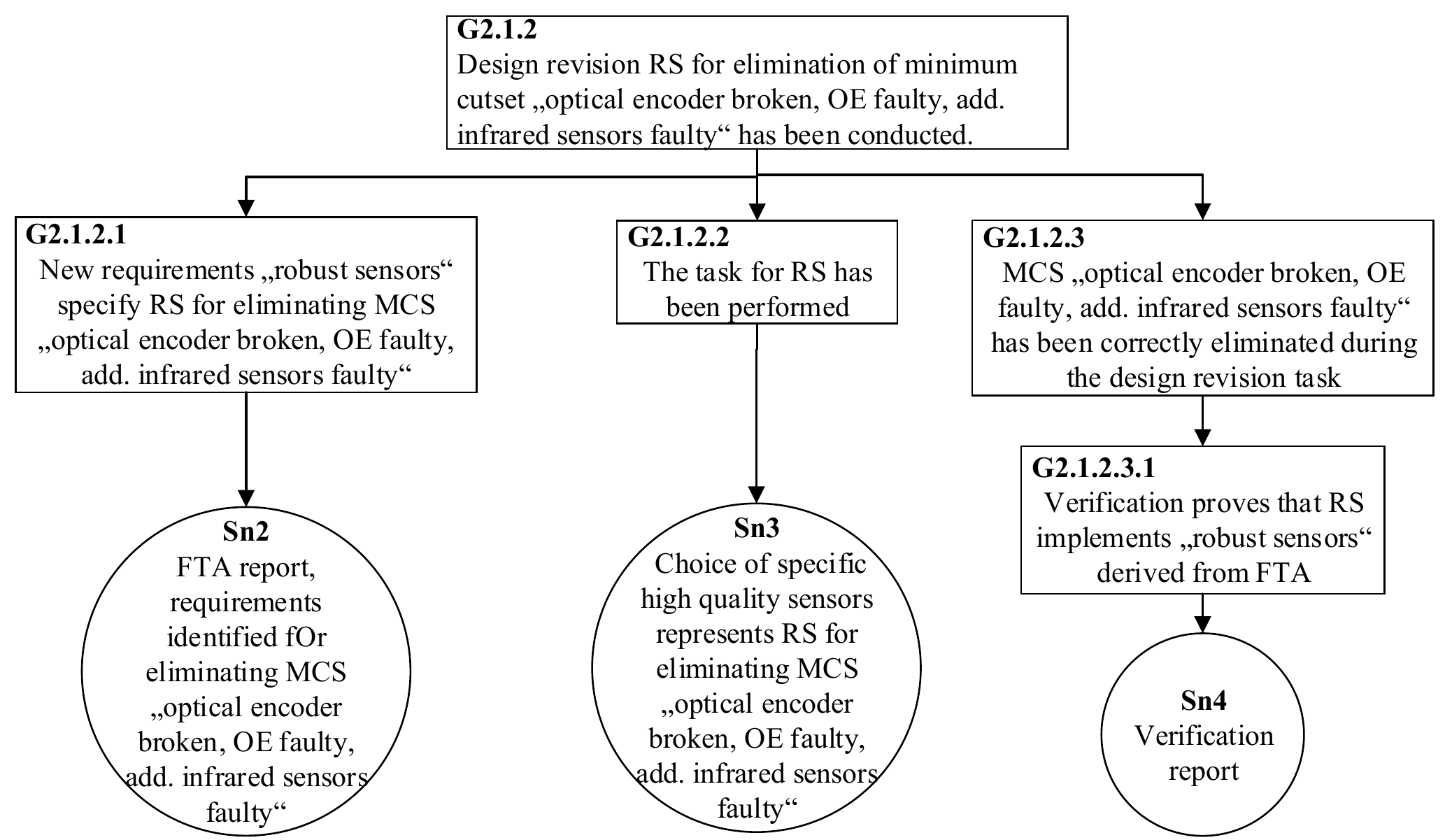}
    \label{fig:pattern:hc:fta:tdcs}
  }\\
  \subfloat[For FMEA]{
    \includegraphics[width=\columnwidth]{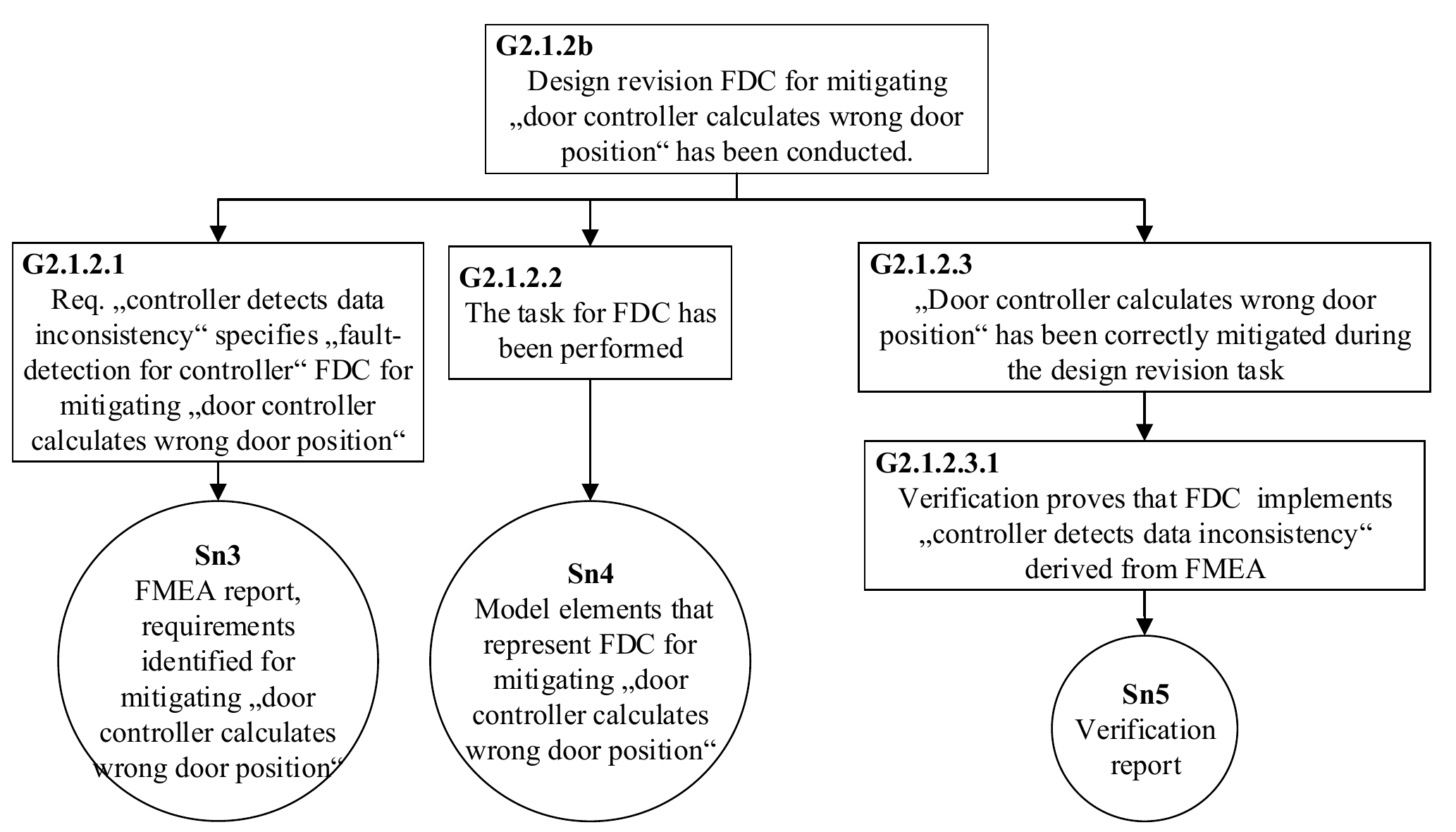}
    \label{fig:pattern:hc:fmea:tdcs}
  }
  \caption{Application of module HC to TDCS}
  \vspace{-.5cm}
\end{figure}

According to \Cref{sec:appl:cr:fta} and
\Cref{fig:pattern:cr:fta:tdcs}, \Cref{fig:pattern:hc:fta:tdcs} shows
the breakdown of Goal 2.1.2 by which we get an argument for
having eliminated this specific critical path.

\Cref{fig:pattern:hc:fmea:tdcs} shows an argument to mitigate the failure
mode identified in \Cref{fig:pattern:cr:fmea:tdcs}: The requirement
``controller detects data inconsistencies'' specifies a task to build
a ``fault-detection for the controller (FDC)'' which is (i) conducted
when we gain evidence for \textbf{Goal 2.1.2.2} by
\textbf{Solution 4}, and (ii) verified as soon as we get evidence for
\textbf{Goal 2.1.2.3} by \textbf{Solution 5}.

\section{Discussion}
\label{sec:disc}

In this section, we reflect on the presented argument pattern and
its application to the TDCS.

\subsection{Preliminary Evaluation in the Student Lab}
\label{sec:prel-eval-stud}

We developed our pattern for a practical course in safety
analysis with 18 master students.\footnote{See
  \url{http://www4.in.tum.de/~gleirsch/safety/index.shtml}.} This
course was combined with an experiment on the comparison of
the effectiveness of FTA, FMEA, and STPA.\footnote{Detailed results
  will be published apart from this paper.} After being 
trained in these techniques, the students worked in groups of three to
perform HA for three different systems, an automotive
anti-lock braking system, an air traffic collision avoidance system,
and a TDCS.

Next, the students (i) developed safety measures for the hazards they
identified by using the three techniques and (ii) constructed a safety
argument. In an extra tutorial on safety cases, we showed them a first
version of our pattern to help them structure their arguments. This
way, we could determine whether trained students were able to apply our
pattern in their assurance tasks. The submissions showed that 4 out of
the 6 groups were able to directly use our pattern to create an
argument from their previous analysis. Finally, this approach helped
us to understand and refine our pattern before evaluating it in a more
critical practical context.

\subsection{Structuring the Argument}
\label{sec:argumentstructure}
Here, we discuss insights from applying our pattern in the course
assignments. From the \emph{construction notes} in
\Cref{sec:patterns}, we conclude that our argument is to be built
top-down.  The structuring is difficult because of the many criteria
and ways available to do this. We found the following
\textbf{classification criteria and steps} helpful to keep the
argument compact:
\begin{enumerate}
\item[(1)] Breakdown of items (\ies system functions, components), %
\item[(2a)] hazard analysis technique, 
\item[(2b)] hazards and measures common across the techniques,
\item[(3a)] requirement types (\ies functional, quality, constraint,
process, see \Cref{sec:concepts}) and clusters, 
\item[(3b)] solution clusters.
\end{enumerate}
For (1), item-related structuring 
\begin{inparaitem}
\item determines analysis granularity in FTA, FMEA, and STPA,
\item structures requirements derived from the identified hazards, and
\item scopes design revisions implementing (functional) requirements.
\end{inparaitem}
For (3), requirements (Goal 2.1.2.1) motivating design revisions
(\egs Goal 2.1.2 in \Cref{fig:pattern:hc:fta:tdcs}) can be clustered
according to their (i) type, and (ii) the
class and severity rating of the hazard they were associated with
(\egs intermittent failure mode in highest RPN range).

We found that these criteria can be used in two sequences:
\[(1) \rightarrow (2) \rightarrow (3)
\quad\mbox{or}\quad
(2) \rightarrow (1) \rightarrow (3).\]

\subsection{Commonalities and Relationships among the Modules}
The modules M and HC have commonalities in their goal
structures. Particularly, the FTA and STPA modules are both deductive
in their causal reasoning which, unsurprisingly, leads to similarities in
their structure.
The mitigation of a system-level event $E$ might be argued from two
different directions: by module CR mitigating a critical path to $E$,
or by module CR mitigating a failure mode having $E$ among its
effects.

\subsection{Evidence for Safety Arguments and Level of Detail}
\label{sec:evidence-detail}
Our pattern is built on evidence from FTA, FMEA, and STPA
results. We refer to evidence on
\begin{inparaitem}
\item \emph{hazards} to argue over their proper identification, and on
\item \emph{countermeasures} to argue over their validity, proper
  implementation, and verification.
\end{inparaitem}

\paragraph*{Semantic Traceability} 
Our modules represent separate concerns facilitating traceability
between complementary evidence (\egs \textit{Why/Is the argument
  complete?}). State-of-the-art patterns pinpoint that hazards have
been mitigated without clearly demonstrating why and how they have
been properly mitigated
\cite{1999_pumfrey_safety_principled_design}. Hence, it is important
to refer to the causal chain of what exactly triggers the hazards and
to how that causal chain is to be modified to eliminate hazard
sources.

\paragraph*{Deductive Argumentation based on Good Practice} 
Gaining confidence in an argument is more of a technical problem,
whereas gaining confidence in the evidence stems from technical,
social and philosophical issues~\cite{2015_rushby_interpretation}. We
focus on increasing confidence in the argument by an optimal
way of constructing it and zooming into (trustworthy) evidence. The
adequacy of evidence itself is out of scope here. Our pattern allows
constructing arguments directly from techniques recommended by
standards, \egs ISO 61508 and 26262.

Our pattern supports deductive argumentation to
reduce doubt. %
Enhanced FTA and FMEA variants might increase the confidence in the
evidence. The pattern employs basic versions of HA techniques,
making it applicable in safety cases of any system, \egs
independent of whether we use enhanced FMEA. 

\paragraph*{Reducing Confirmation Bias}
\label{sec:addr-conf-bias}
We address the problem of confirmation bias as the ``tendency for
people to favor information that confirms their preconceptions or
hypotheses, regardless of whether the information is true''
\cite{Leveson2011}. This bias due to the goal ``to show that the
system is safe'' is reduced because previously conducted FTA, FMEA,
and STPA use tactics to collect evidence for the goal ``to show that
there are hazards.'' This way, our pattern supports two-staged
arguments, the first stage to be constructed already during the design
stage as required by, \egs Leveson~\cite{Leveson2011} and Yuan and
Xu~\cite{Yuan2010}.

\paragraph*{Using Specific Terminology} 
Our pattern contains claims based on HA terminology. Any
person reviewing the safety case has to know HA. However,
the module structure supports exploring technical details,
it is complementary to existing patterns, and helps strengthen 
arguments to be assessed by certification engineers.

\subsection{Applicability, Soundness, and Relative Completeness}
We informally investigate three criteria to
argue for the usability of the discussed pattern:

\paragraph*{Applicability}
\label{sec:applicability}
Hawkins et al.~\cite{Hawkins2011} offer attributes against which
argument quality can be scrutinized. Based on
\Cref{sec:prel-eval-stud}, we believe that our pattern is (i)
\textit{easy to understand and apply by software engineers} and (ii)
\textit{flexible enough to be applicable to many safety-critical
  systems}, as it has been applied to three control
systems in fairly different domains by a group of 18 students.

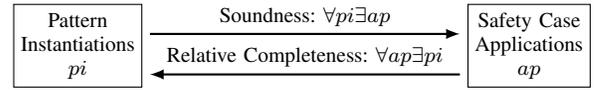
\begin{figure}
  \centering
  \footnotesize
\begin{tikzpicture}
  [box/.style={draw,align=center,minimum height=1.1cm},
  rstage/.style={align=left}]
  \node[box] (patinst) at (0,0) {Pattern\\Instantiations\\$pi$};
  \node[above=0.2em of patinst.east] (p1) {}; 
  \node[below=1em of patinst.east] (p2) {}; 
  \node[box] (scappl) at (6,0) {Safety Case\\Applications\\$ap$};
  \node[above=0.2em of scappl.west] (s1) {}; 
  \node[below=1em of scappl.west] (s2) {}; 
  \draw[arrows={-latex[length=10pt]},thick] 
  (p1) edge[above] node {Soundness: $\forall pi \exists ap$} (s1)
  (s2) edge[above,align=center] node {Relative Completeness: $\forall ap \exists pi$} (p2);
\end{tikzpicture}
   \caption{Soundness and relative completeness}
  \label{fig:soundcomplete}
  \vspace{-.5cm}
\end{figure}

\paragraph*{Soundness}
\label{sec:soundness}
\emph{Does any instantiation of the pattern form a sound safety
  argument based on FTA, FMEA, and STPA?}  First, the module $CR$
resembles not only the causal reasoning direction, but can also
directly use any result of these analyses, \ies any identified
hazard. Second, module $HC$ provides a response to this hazard in
terms of an identified and taken countermeasure whose verification is
part of $HC$. A further discussion of this question is out of scope
here (\Cref{fig:soundcomplete}).

\paragraph*{Relative Completeness}
\label{sec:completeness}
\emph{Can the pattern be instantiated to the most relevant situations
  where safety cases are based on FTA, FMEA, and STPA?}  Here, we
elaborate on applicability aspect (ii): The described modules capture
core concepts of the three techniques, such that to each case
where one of these techniques is applicable we can also expect to be
able to instantiate our pattern (\Cref{fig:soundcomplete}).

\paragraph*{Assessing Hazard Analysis Techniques}
Some principles to be followed by good techniques
\cite{1999_pumfrey_safety_principled_design} are
embedded in our pattern:
\textit{Method is more important than notation} emphasizes clear
description of the capabilities of a technique, specification of
information sources and the analysis procedure.
\textit{Techniques should use familiar concepts and models} implies
that trustworthy results should be derived from HA, using combinations
of events and conditions to model causes and effects. Our pattern
incorporates HA steps (see \Cref{sec:evidence-detail}). Trying to
use our pattern with a new HA technique may unveil problems if
this technique violates this principle. By refining the modules for a
new technique, we can assess whether this technique matches at
least as good as the existing ones.

\section{Related Work}
\label{sec:relwork}

Alexander et al.~\cite{Alexander2007} present patterns arguing for
safe control software by using adaptation mechanisms for improving or
maintaining safe states.  Our three modules (M, CR, and HC) are
similar to their core patterns but use fewer argumentation steps.
They focus on FMEA, \egs arguing over adaptation as a
design measure ($DR$) which goes beyond \emph{Solution 4} in our
countermeasure module ($HC$).  We integrate FTA, FMEA, and STPA and
capture how safety requirements were motivated.  We aim at a
compact pattern and use HA results more
directly.  Their argument depicts that an identified risk structure
(\ies failure modes as hazards and their causal factors) is acceptable
and, by adaptations as countermeasures, how this structure
got acceptable.  We do not presume adaptation as a
countermeasure. Moreover, we allow the goal structure to be built
during FTA, FMEA, or STPA %
whereas they concentrate on an a-posteriori construction.

Kelly~\cite[pp.~317ff]{Kelly1998} and the GSN standard~\cite{GSN2011}
describe a ``fault tree evidence'' pattern where a fault tree as a
whole serves as an evidence to derive basic safety goals. We go more
in-depth into fault tree analysis and make the argument more precise.
Similar to an application of our causal reasoning module ($CR$) for
FTA, \cite[p.~76f]{Kelly1998} describes an argument that includes
cutsets at the level of solutions in the goal structure, however, not
elaborated as a pattern.

Hawkins and Kelly~\cite{Hawkins2013} present patterns for mitigation
of software-based hazards, identification and realization of software
safety requirements, and avoidance of software-based mistakes.  While
their ``software contribution safety argument'' module %
is similar to our HC module, they do not elaborate on causal reasoning
based on a specific HA and mitigation technique. Their article does
not discuss system models.

Palin and Habli~\cite{2010_palin_automotive_assurance} propose safety
case pattern catalogs for the construction of a vehicle 
safety case in accordance with ISO~26262. One
of the proposed catalogs, namely the \textit{Architecture for the
  Argument} pattern catalog, confirms the necessity of an
\textit{FMEA Argument} pattern. Their paper does not elaborate
on this pattern. 

For a truck information and control system, Dardar~\cite{Dardar2014}
constructs an ISO~26262-compliant safety case using GSN and SysML. He
argues from a trustworthy process based on coarse evidence from
several HA and quality assurance techniques.
However, for FTA he shows a goal structure that can be seen as an
instance of our CR and HC modules for FTA.

Wagner~\cite{Wagner2012} sketches an argument pattern based on
STPA. Our CR module for STPA indicates more clearly how STPA results
can be integrated into the argument.  Beyond our M, CR, and HC
modules, Wagner proposes modules to capture process- and
environment-based arguments~\cite{Wagner2010}.

Research has been done on assurance deficits due to limitations of FTA
and FMEA~\cite{2012_shebl_failure, 2012_sun_confidence_phd}. Our goal
is not to tackle these problems, but to construct confident arguments
in case of proper HA. However, these deficits need to be addressed.

\section{Conclusions and Future Work}
\label{sec:conc}

We presented argumentation from the contribution of HA techniques to a
system's safety generalized by a modular argument pattern.  
We showed by an example and by discussion that this pattern (i)
captures the structure of these causal reasoning techniques and (ii)
extracts commonalities in reasoning and evidence among the three
techniques.
Furthermore, we broke down the evidence argument based on the
solutions by using the causal reasoning and mitigation structure
coming with the HA and prepared our pattern to be integrated with
system models.
Next, we asked trained master students in a practical course to apply
a preliminary version of our pattern to construct their safety cases
of real-life applications.
Finally, we added value to HA by (i) integrating results scattered
across several specialty HA techniques and (ii) integrating these
results with the additional steps required in the construction of
balanced, complete, and confident safety cases.

\paragraph*{Future Work}
After having interviewed our students,
we can evaluate the usefulness of our pattern in an
experiment with safety compliance practitioners.
Confidence is increased by the fact that the argument structure
mirrors the steps and the causal reasoning of an HA
technique. Elaborating on arguing that an application of an HA
technique is trustworthy is, however, an important direction to be
investigated. Moreover, we plan to work on a formalization for an
automated pattern instantiation.
Finally, we consider a manual for using our pattern as important
for the transfer to practice.

\paragraph*{Acknowledgments}
We thank our students of the advanced practical course for applying
the argument pattern in their course assignments.

 \bibliography{assurancepatterns}

\begin{thebibliography}{10}
\providecommand{\url}[1]{#1}
\csname url@samestyle\endcsname
\providecommand{\newblock}{\relax}
\providecommand{\bibinfo}[2]{#2}
\providecommand{\BIBentrySTDinterwordspacing}{\spaceskip=0pt\relax}
\providecommand{\BIBentryALTinterwordstretchfactor}{4}
\providecommand{\BIBentryALTinterwordspacing}{\spaceskip=\fontdimen2\font plus
\BIBentryALTinterwordstretchfactor\fontdimen3\font minus
  \fontdimen4\font\relax}
\providecommand{\BIBforeignlanguage}[2]{{%
\expandafter\ifx\csname l@#1\endcsname\relax
\typeout{** WARNING: IEEEtran.bst: No hyphenation pattern has been}%
\typeout{** loaded for the language `#1'. Using the pattern for}%
\typeout{** the default language instead.}%
\else
\language=\csname l@#1\endcsname
\fi
#2}}
\providecommand{\BIBdecl}{\relax}
\BIBdecl

\bibitem{Bishop1998}
P.~Bishop and R.~Bloomfield, ``A methodology for safety case development,'' in
  \emph{6th Safety-critical Systems Symp.}, F.~Redmill and T.~Anderson,
  Eds.\hskip 1em plus 0.5em minus 0.4em\relax Birmingham, UK: Springer, Feb
  1998.

\bibitem{Kelly1998}
T.~P. Kelly, ``Arguing safety -- a systematic approach to safety case
  management,'' Ph.D. dissertation, Dept. of Comp. Sci., University of York,
  UK, Sep 1998.

\bibitem{Kelly1997}
T.~P. Kelly and J.~A. McDermid, ``Safety case construction and reuse using
  patterns,'' in \emph{Computer Safety, Reliability and Security: SAFECOMP 97.
  16th International Conference}, P.~Daniel, Ed.\hskip 1em plus 0.5em minus
  0.4em\relax London: Springer, 1997, pp. 55--69.

\bibitem{Alexander2007}
R.~Alexander, T.~P. Kelly, Z.~Kurd, and J.~A. McDermid, ``Safety cases for
  advanced control software: Safety case patterns,'' DTIC Document, Tech. Rep.,
  2007.

\bibitem{Rausand1996}
M.~Rausand and K.~{\O}ien, ``The basic concepts of failure analysis,''
  \emph{Reliability Engineering \& System Safety}, vol.~53, no.~1, pp. 73--83,
  1996.

\bibitem{Gleirscher2014a}
M.~Gleirscher, ``Behavioral safety of technical systems,'' Dissertation,
  Technische Universit{\"a}t M{\"u}nchen, Dec. 2014.

\bibitem{Chillarege1992}
R.~Chillarege, I.~Bhandari, J.~Chaar, M.~Halliday, D.~Moebus, B.~Ray, and
  M.~Wong, ``Orthogonal defect classification -- a concept for in-process
  measurements,'' \emph{IEEE TASE}, vol.~18, no.~11, pp. 943--56, 1992.

\bibitem{Leveson2004}
N.~G. Leveson, ``A new accident model for engineering safer systems,''
  \emph{Safety Science}, vol.~42, no.~4, pp. 237--70, 2004.

\bibitem{Beizer1990}
B.~Beizer, \emph{Software Testing Techniques}, 2nd~ed.\hskip 1em plus 0.5em
  minus 0.4em\relax New York: Van Nostrand Reinhold, Jun 1990.

\bibitem{Ericson2015}
C.~A. Ericson, \emph{Hazard Analysis Techniques for System Safety},
  2nd~ed.\hskip 1em plus 0.5em minus 0.4em\relax Wiley, 7 2015.

\bibitem{Leveson2012}
N.~G. Leveson, \emph{Engineering a Safer World: Systems Thinking Applied to
  Safety}, ser. Engineering Systems.\hskip 1em plus 0.5em minus 0.4em\relax MIT
  Press, Jan 2012.

\bibitem{Fenelon1994}
P.~Fenelon, J.~A. McDermid, M.~Nicolson, and D.~J. Pumfrey, ``Towards
  integrated safety analysis and design,'' \emph{SIGAPP Appl. Comput. Rev.},
  vol.~2, no.~1, pp. 21--32, Mar 1994.

\bibitem{Hawkins2015}
R.~Hawkins, I.~Habli, D.~Kolovos, R.~Paige, and T.~Kelly, ``Weaving an
  assurance case from design: A model-based approach,'' in \emph{16th IEEE
  International Symposium on High Assurance Systems Engineering}, Jan 2015, pp.
  110--117.

\bibitem{Wassyng2010}
A.~Wassyng, T.~S.~E. Maibaum, M.~Lawford, and H.~Bherer, ``Software
  certification: Is there a case against safety cases?'' in \emph{Foundations
  of Computer Software. Modeling, Development, and Verification of Adaptive
  Systems - 16th Monterey Workshop}, 2010, pp. 206--27.

\bibitem{Weinstock2013}
C.~Weinstock, J.~Goodenough, and A.~Klein, ``Measuring assurance case
  confidence using \textsc{Baconian} probabilities,'' in \emph{Assurance Cases
  for Software-Intensive Systems (ASSURE), 2013 1st International Workshop on},
  May 2013, pp. 7--11.

\bibitem{2015_graydon_epistemology}
P.~Graydon and C.~Holloway, ``Evidence under a magnifying glass: Thoughts on
  safety argument epistemology,'' \emph{IET Conf Proc}, 2015.

\bibitem{Dardar2014}
R.~Dardar, ``Building a safety case in compliance with iso 26262 for fuel
  levelestimation and display system,'' Master's thesis, M\"alardalen
  University, School of Innovation, Design and Engineering, 2014.

\bibitem{Yuan2010}
T.~Yuan and T.~Xu, ``Computer system safety argument schemes,'' in
  \emph{Software Engineering (WCSE), 2010. 2nd World Congress on}, vol.~2, Dec
  2010, pp. 107--110.

\bibitem{GSN2011}
\emph{{GSN} Community Standard}, Origin Consulting Ltd Std., Rev. VERSION 1,
  Nov. 2011.

\bibitem{Leveson2011}
N.~Leveson, ``The use of safety cases in certification and regulation,''
  \emph{Journal of System Safety}, vol.~47, no.~6, pp. e--Edition, 2011.

\bibitem{1999_pumfrey_safety_principled_design}
D.~J. Pumfrey, ``The principled design of computer system safety analyses.''
  Ph.D. dissertation, University of York, 1999.

\bibitem{2015_rushby_interpretation}
J.~Rushby, ``The interpretation and evaluation of assurance cases,'' Technical
  report SRI-CSL-15-01, Computer Science Laboratory, SRI International, Menlo
  Park, CA, Tech. Rep., 2015.

\bibitem{Hawkins2011}
R.~Hawkins, K.~Clegg, R.~Alexander, and T.~Kelly, ``Using a software safety
  argument pattern catalogue: Two case studies,'' in \emph{30th
  SAFECOMP}.\hskip 1em plus 0.5em minus 0.4em\relax Berlin, Heidelberg:
  Springer, 2011, pp. 185--98.

\bibitem{Hawkins2013}
R.~Hawkins and T.~Kelly, ``A software safety argument pattern catalogue,'' The
  University of York, Tech. Rep. YCS-2013-482, 2013.

\bibitem{2010_palin_automotive_assurance}
R.~Palin and I.~Habli, ``Assurance of automotive safety--a safety case
  approach,'' in \emph{International Conference on Computer Safety,
  Reliability, and Security}, 2010.

\bibitem{Wagner2012}
\BIBentryALTinterwordspacing
S.~Wagner, ``Combining {STAMP/STPA} and assurance cases,'' 2012, slides only.
  [Online]. Available:
  \url{http://psas.scripts.mit.edu/home/get_pdf.php?name=3-4-Wagner-Combining-STPA-and-Assurance-Cases.pdf}
\BIBentrySTDinterwordspacing

\bibitem{Wagner2010}
S.~Wagner, B.~Sch{\"a}tz, S.~Puchner, and P.~Kock, ``A case study on safety
  cases in the automotive domain: Modules, patterns, and models,'' in
  \emph{21st Int. Symp. Software Reliability Engineering}.\hskip 1em plus 0.5em
  minus 0.4em\relax IEEE, 2010, pp. 269--78.

\bibitem{2012_shebl_failure}
N.~A. Shebl, B.~D. Franklin, and N.~Barber, ``Failure mode and effects analysis
  outputs: are they valid?'' \emph{BMC health services research}, 2012.

\bibitem{2012_sun_confidence_phd}
L.~Sun, ``Establishing confidence in safety assessment evidence,'' Ph.D.
  dissertation, The University of York, 2012.

\end{thebibliography}
\end{document}